\documentclass[12pt,preprint]{emulateapj}

\def\H0{{\rm ~km~s^{-1}~Mpc^{-1}}}

\def\go{
\mathrel{\raise.3ex\hbox{$>$}\mkern-14mu\lower0.6ex\hbox{$\sim$}}
}
\def\lo{
\mathrel{\raise.3ex\hbox{$<$}\mkern-14mu\lower0.6ex\hbox{$\sim$}}
}

\shorttitle{GRB 061121: Broadband observations}
\shortauthors{K.L. Page et al.}

\begin{document}

\title{GRB 061121: Broadband spectral
  evolution through the prompt and afterglow phases of a bright burst.}

\author{K.L. Page\altaffilmark{1}, R. Willingale\altaffilmark{1},
  J.P. Osborne\altaffilmark{1}, B. Zhang\altaffilmark{2},
  O. Godet\altaffilmark{1}, F.E. Marshall\altaffilmark{3}, A. Melandri\altaffilmark{4},
  J.P. Norris\altaffilmark{5,6}, P.T. O'Brien\altaffilmark{1},
   V. Pal'shin\altaffilmark{7},
  E. Rol\altaffilmark{1}, P. Romano\altaffilmark{8,9},
  R.L.C. Starling\altaffilmark{1}, P. Schady\altaffilmark{10},
  S.A. Yost\altaffilmark{11}, 
  S.D. Barthelmy\altaffilmark{3}, A.P. Beardmore\altaffilmark{1}, G. Cusumano\altaffilmark{12},
   D.N. Burrows\altaffilmark{13}, M. De Pasquale\altaffilmark{10},
  M. Ehle\altaffilmark{14}, P.A. Evans\altaffilmark{1}, 
  N. Gehrels\altaffilmark{3}, M.R. Goad\altaffilmark{1}, S. Golenetskii\altaffilmark{7},
  C. Guidorzi\altaffilmark{8,9},  C. Mundell\altaffilmark{4},
   M.J. Page\altaffilmark{10}, G. Ricker\altaffilmark{15},
   T. Sakamoto\altaffilmark{3},  B.E. Schaefer\altaffilmark{16}, M. Stamatikos\altaffilmark{3},
    E. Troja\altaffilmark{1,12}, M.Ulanov\altaffilmark{7}, F. Yuan\altaffilmark{11} \&
   H. Ziaeepour\altaffilmark{9}}

\altaffiltext{1}{Department of Physics and Astronomy, University of Leicester,
  Leicester, LE1 7RH, UK}
\altaffiltext{2}{Department of Physics \& Astronomy, University of Nevada, Las
  Vegas, NV 89154-4002, USA}
\altaffiltext{3}{NASA/Goddard Space Flight
  Center, Greenbelt, MD 20771, USA}
\altaffiltext{4}{Astrophysics Research Institute, Liverpool John Moores University, Twelve
Quays House, Egerton Wharf, Birkenhead, CH41 1LD}
\altaffiltext{5}{Denver Research Institute, University of Denver, Denver, CO
  80208, USA}
\altaffiltext{6}{Visiting Scholar, Stanford University}
\altaffiltext{7}{Ioffe Physico-Technical Institute, Laboratory for
  Experimental Astrophysics, 26 Polytekhnicheskaya, Saint Petersburg 194021,
  Russian Federation}
\altaffiltext{8}{INAF, Osservatorio Astronomico di Brera, Via E. Bianchi 46,
I-23807, Merate (LC), Italy}
\altaffiltext{9}{Dipartimento di Fisica, Universit{\'a} di Milano-Bicocca,
  Piazza delle Scienze 3, I-20126, Milano, Italy}
\altaffiltext{10}{Mullard Space Science Laboratory, University College London, Holmbury St. Mary, Dorking, Surrey RH5 6NT, UK}
\altaffiltext{11}{University of Michigan, 2477 Randall Laboratory, 450 Church
  St., Ann Arbor, MI 48104, USA}
\altaffiltext{12}{INAF-IASF, Sezione di Palermo, via Ugo La Malfa 153, 90146,
  Palermo, Italy}
\altaffiltext{13}{Department of Astronomy and Astrophysics, Pennsylvania State
  University, 525 Davey Lab, University Park, PA 16802, USA}
\altaffiltext{14}{XMM-Newton Science Operations Centre, European Space Agency,
  Villafranca del Castillo, Apartado 50727, E-28080 Madrid, Spain}
\altaffiltext{15}{Center for Space Research, Massachusetts Institute of
  Technology, 70 Vassar Street, Cambridge, MA 02139, USA}
\altaffiltext{16}{Department of Physics and Astronomy, Louisiana State
  University, Baton Rouge, LA 70803, USA}

\email{kpa@star.le.ac.uk}

\begin{abstract}

{\it Swift} triggered on a precursor to the main burst of GRB~061121 (z~=~1.314),
allowing observations to be made from the optical to gamma-ray bands. Many other telescopes, including {\it Konus-Wind}, {\it XMM-Newton}, {\it
  ROTSE} and the Faulkes Telescope North, also observed the burst.
The
gamma-ray, X-ray and UV/optical emission all showed a peak $\sim$~75~s after
the trigger, although the optical and X-ray afterglow components also appear early
on -- before, or during, the main peak. 
Spectral evolution was seen
throughout the burst, with the prompt emission showing a clear positive correlation between
brightness and hardness. 
The Spectral Energy Distribution (SED) of the prompt emission, stretching from 1~eV up
to 1~MeV, is very flat, with a peak in the flux density at $\sim$~1~keV. The
optical-to-X-ray spectra at this time are better fitted by a broken, rather
than single, power-law, similar to previous results for X-ray flares. The
SED shows spectral hardening as the afterglow evolves
with time. This
behaviour might be a symptom of self-Comptonisation, although
circumstellar densities similar to those found in the cores of molecular
clouds would be required. The afterglow also decays too
slowly to be accounted for by the standard models.
Although the
precursor and main emission show different spectral lags, both are consistent
with the lag-luminosity correlation for long bursts. 
GRB~061121 is the instantaneously brightest long burst yet detected by {\it
  Swift}. 
Using a combination of
{\it Swift} and {\it Konus-Wind} data, we estimate an isotropic energy of
2.8~$\times$~10$^{53}$~erg over 1~keV -- 10~MeV in the GRB rest frame. A
probable jet break is detected at $\sim$~2~$\times$~10$^{5}$~s, leading to an
estimate of $\sim$~10$^{51}$~erg for the beaming-corrected gamma-ray energy.

\end{abstract}


\section{Introduction}

Gamma-Ray Bursts (GRBs) are intrinsically extremely luminous objects, approaching
values of 10$^{54}$ erg~s$^{-1}$ if the radiation is isotropic (e.g., Frail et
al. 2001; Bloom et al.\ 2003). 
This energy is emitted over all bands in the electromagnetic spectrum; to
understand GRBs as fully as possible, panchromatic observations are required over all
time frames of the burst.

The {\it Swift} multi-wavelength observatory (Gehrels et al.\ 2004) is designed
to detect and follow-up GRBs. 
With its rapid slewing ability, {\it Swift} is able to follow bursts and their afterglows
from less than a minute after the initial trigger, and can often still detect
them weeks, and sometimes months, later. On rare occasions,
such as when {\it Swift} triggers on a precursor to the main burst,
the prompt emission, as well as the afterglow, can be observed at X-ray and
UV/optical wavelengths. GRB~061121, the subject of this paper,
is only the third GRB {\it Swift} has detected in this manner (after
GRB~050117 -- Hill et al.\ 2006 and GRB~060124 -- Romano et al.\ 2006), out of
the almost 200 bursts triggered on in the first two years of the mission.\footnote{GRB~050820A would possibly have
also been in this category, but {\it Swift} entered the South Atlantic Anomaly (SAA)
just as a dramatic increase in count rate began (Cenko et al.\ 2006;
Page et al.\ 2005a;
Cummings et al.\ 2005; Page et al.\ 2005b; Chester et al.\ 2005); {\it Swift}
does not actively collect data during SAA passages.} Of these, GRB~061121 is the second well-sampled event (GRB~060124 was the first), and the first for which the UV/Optical
Telescope (UVOT) was in event mode.

In addition to the small number of precursor triggers, around 10\% of {\it
  Swift} bursts show detectable emission over the BAT bandpass by the time the narrow field instruments (NFIs) are on target.

Besides the {\it Swift} observations of prompt emission, there have been a
small number of prompt optical measurements of GRBs, thanks to the
increasing number of robotic telescopes around the world. A variety of
behaviours has been found, with some optical (and infrared) light-curves tracking
the gamma-ray emission (e.g., GRB~041219A -- Vestrand et al.\ 2005; Blake et
al.\ 2005), while others appear uncorrelated (e.g., GRB~990123 -- Akerlof et
al.\ 1999, Panaitescu \& Kumar 2007, though see also Tang \& Zhang 2006; GRB~050904 -- Bo{\" e}r et al.\ 2006;
GRB~060111B -- Klotz et al.\ 2006; GRB~060124 -- Romano et
al.\ 2006). GRB~050820A (Vestrand et al.\ 2006) showed a
mixture of both correlated and uncorrelated optical flux.

Where correlations exist between different energy bands, it
is likely that there is a common origin for the components. In the
uncorrelated cases, the optical emission may be due to an external reverse shock (e.g., Sari \& Piran 1999;
M{\' e}sz{\' a}ros \& Rees 1999), while the prompt gamma-rays are caused by
internal shocks. Cenko et al.\ (2006) suggest that the early optical data for
GRB~050820A are produced by the forward shock passing through the band.
In the case of GRB~990123, Panaitescu \& Kumar (2007) have
suggested that the gamma-rays arose from inverse Comptonisation, while the
optical emission was due to synchrotron processes; they do not assume a
specific mechanism for the energy dissipation, allowing for the possibility of
either internal or reverse-external shocks.

It is unclear whether precursors are ubiquitous features of GRBs, often remaining
undetected because of a low signal-to-noise ratio or being outside the
energy bandpass of the detector, or whether only some bursts exhibit them. A
detailed discussion of the precursor phenomenon is beyond the scope of this paper
and will be addressed in a future publication.

In this paper, we report on the multi-wavelength observations of both the
prompt and afterglow emission of GRB~061121. $\S$\ref{obs} details the
observations made by {\it Swift}, {\it Konus-Wind}, {\it XMM-Newton}, {\it
  ROTSE}\footnote{Robotic Optical Transient Search Experiment} and the Faulkes Telescope North (FTN), with multi-band
comparisons being made. In $\S$\ref{disc},
we discuss the precursor, prompt and afterglow emission, with a summary given in $\S$\ref{conc}.

Throughout the paper, the main burst ($\sim$~60--200~s
after the trigger) will be referred to as the prompt emission, and the
emission seen over $-$5 to +10~s as the precursor, where the BAT trigger time T$_{0}$~=~0~s. Errors are given at 90\%
confidence (e.g., $\Delta\chi^{2}$~=~2.7 for one interesting parameter) unless
otherwise stated, and the convention F$_{\rm
  \nu,t}$~$\propto$~$\nu^{-\beta}$t$^{-\alpha}$ 
(with the photon spectral index, $\Gamma$~=~$\beta$~+~1 where dN/dE~$\propto$~E$^{-\Gamma}$) has been
followed. We have assumed  a flat
Universe, with Hubble constant,
H$_{0}$~=~70~km~s$^{-1}$~Mpc$^{-1}$, cosmological constant,
$\Omega_{\Lambda}$~=~0.73 and $\Omega_{\rm matter}$~=~1$-$$\Omega_{\Lambda}$.

\section{Observations and Analyses}
\label{obs}

Two years and one day after launch, the Burst Alert Telescope (BAT; Barthelmy
et al.\ 2005) triggered on a precursor to GRB~061121 at
15:22:29~UT on 21st November, 2006. {\it Swift} slewed immediately, resulting
in the NFIs
 being on target and beginning to collect data 55~s (X-ray
Telescope: XRT; Burrows et al.\ 2005a) and 62~s (UVOT;
Roming et al.\ 2005)
later. This enabled broadband observations of the main burst event, which
peaked $\sim$~75~s after the trigger, leading to spectacular multi-wavelength coverage
of the prompt emission. The most accurate {\it Swift} position for this burst
was that determined by the UVOT: RA~=~09$^{\rm h}$~48$^{\rm m}$~54$\fs$55,
decl~=~$-$13$^{\circ}$~11$^{\prime}$~42$\farcs$4 (J2000.0; 90\% confidence
radius of 0$\farcs$6; Marshall et al.\ 2006); the refined XRT position is only 0$\farcs$1 from
these coordinates (Page et al.\ 2006b).

GRB~061121 was declared a `burst of interest' by the
{\it Swift} team (Gehrels et al.\ 2006a), to encourage an intensive ground-
and space-based follow-up
programme. In addition to the {\it Swift} observations, the prompt emission of
GRB~061121 was detected by {\it RHESSI}\footnote{Reuven Ramaty High Energy Solar Spectroscopic Imager} (Bellm et al.\ 2006), {\it
  Konus-Wind} and {\it  Konus-A} (Golenetskii et al.\ 2006). Later afterglow
observations were obtained in the X-ray ({\it XMM-Newton} -- Schartel 2006) and radio 
  (VLA\footnote{Very Large Array} -- Chandra \& Frail 2006) bands. ATCA\footnote{Australia Telescope Compact Array} and WSRT\footnote{Westerbork Synthesis Radio Telescope} also observed in
  the radio band between $\sim$5.2~day and $\sim$6.2~day after the burst, but did not detect the afterglow (van der Horst et al.\ 2006a,b), implying it had faded since the VLA observation.

Likewise, extensive optical follow-up observations were performed: {\it ROTSE-IIIa} (Yost et al.\ 2006), FTN (Melandri et al.\ 2006), Kanata
  1.5-m telescope (Uemura et al.\ 2006), the University of Miyazaki
  30-cm telescope (Sonoda et al.\ 2006), MDM\footnote{Michigan-Dartmouth-MIT
  Observatory} (Halpern et al.\ 2006a,b; Halpern \& Armstrong 2006a,b),
  P60\footnote{Palomar 60 inch}
  (Cenko 2006), ART\footnote{Automated Response Telescope}
  (Torii 2006), the CrAO\footnote{Crimean Astrophysical Observatory} 2.6-m telescope
  (Efimov et al.\ 2006a,b) and SMARTS/ANDICAM\footnote{Small and Moderate Aperture Research Telescope System/A Novel Double-Imaging CAMera} (at
  infrared wavelengths, too; Cobb 2006) all detected the optical
  afterglow. Spectroscopic observations were performed at the Keck telescope
  about 12~minute after the trigger, finding a redshift of
  $z$~=~1.314 for the optical afterglow, based on absorption features (Perley \& Bloom 2006; Bloom et al.\ 2006).

GRB~061121 has the highest instantaneous peak flux of all the long bursts detected by {\it
  Swift} to date (e.g., Angelini et
al.\, in prep).

\subsection{Gamma-ray Data}

\subsubsection{BAT}

\paragraph{Temporal Analysis}

After the initial precursor, the BAT count
rate returned to close to the instrumental background level, until
T$_{0}$+60~s, at which point the much brighter main burst began. This is
characterised by a series of overlapping peaks, each brighter than the
previous one, after which the gamma-ray flux decayed (from
$\sim$T$_{0}$+75~s to $\sim$T$_{0}$+140~s). Event data were
collected until almost 1~ks
after the trigger, thus covering the entire emission period. 

\begin{figure*}
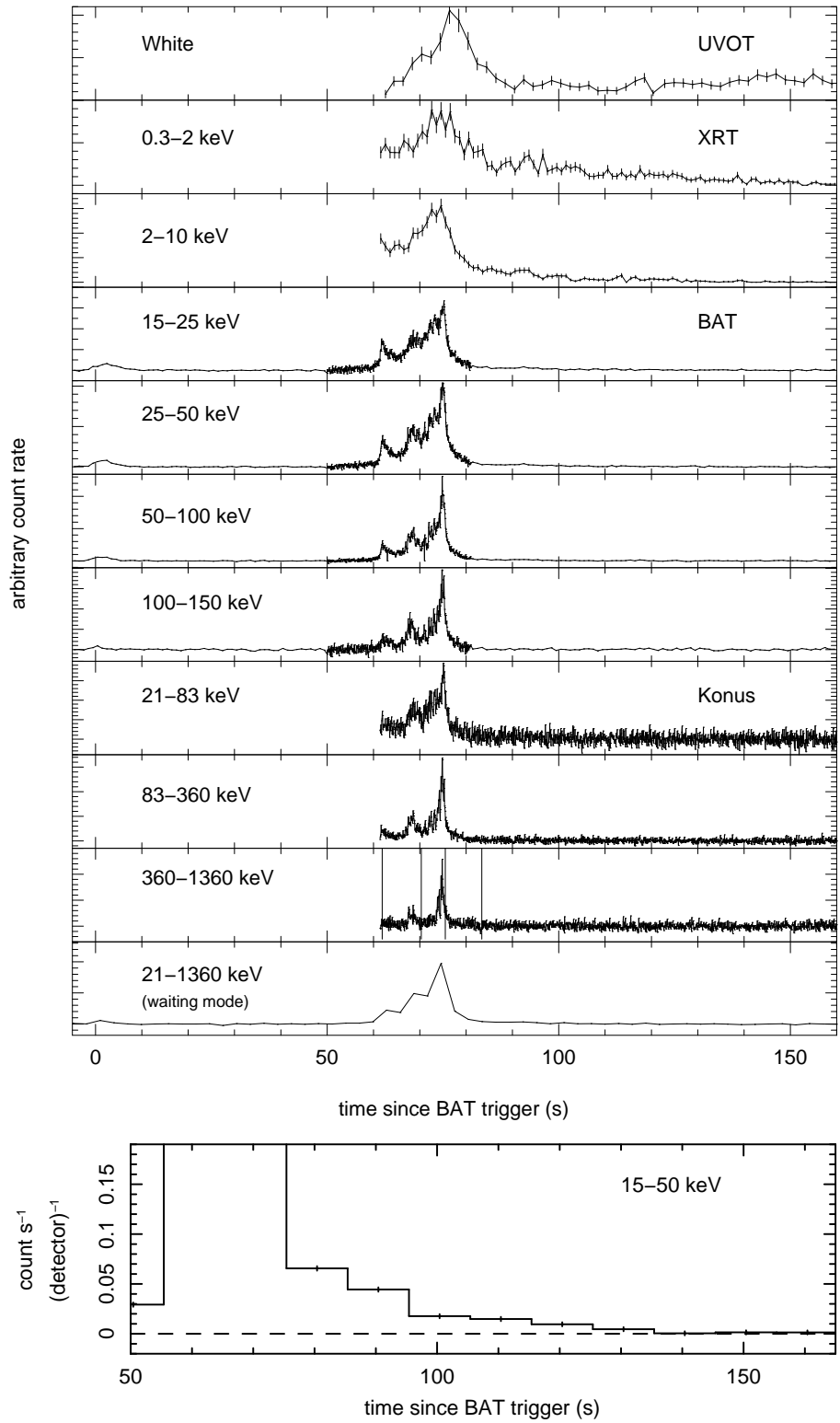

\begin{center}
\includegraphics[clip,width=12.0cm]{f1a.ps}
\includegraphics[clip,angle=-90, width=12.0cm]{f1b.ps}
\caption{Top panels: {\it Swift} UVOT, XRT, BAT and {\it Konus-Wind} light-curves of
  GRB~061121; 1$\sigma$ error bars are shown for the UVOT and XRT data. Each instrument detected the peak of the main burst, with the
  precursor being detected over all gamma-ray energies. The vertical lines in
  the 360--1360~keV panel indicate the start and stop times for the spectra
  given in Table~\ref{kwtab}. Bottom panel: The 15-50~keV BAT light-curve,
  with 10-s bins, showing a tail out to $\sim$140~s.}
\label{allbandlc}
\end{center}
\end{figure*}

T$_{90}$, over 15-150 keV, and incorporating both the
precursor and main emission, is 81~$\pm$~5~s, measured from 8.8--89.8~s after the trigger\footnote{Errors on the BAT T$_{90}$ are
estimated to be typically 5--10\%, depending on the shape of the
light-curve.}. 
Figure~\ref{allbandlc} shows the mask-weighted BAT light-curve in the four standard energy
bands [15--25, 25--50, 50--100, 100-150~keV; 64~ms binning between 50-80~s
after the trigger, with 1~s bins at all other times; units of count~s$^{-1}$~(fully illuminated detector)$^{-1}$], with light-curves
from other instruments: the precursor and the pulses of
the main burst are detected over all gamma-ray bands, although the precursor
is only marginal over the 100-150~keV BAT band. There is also a soft tail
(detected below $\sim$~50~keV, when sufficiently coarse time bins are used) visible until about 140~s after the trigger
(see bottom panel of Figure~\ref{allbandlc}), corresponding to a similar
feature in the X-ray light-curves.


\begin{figure}
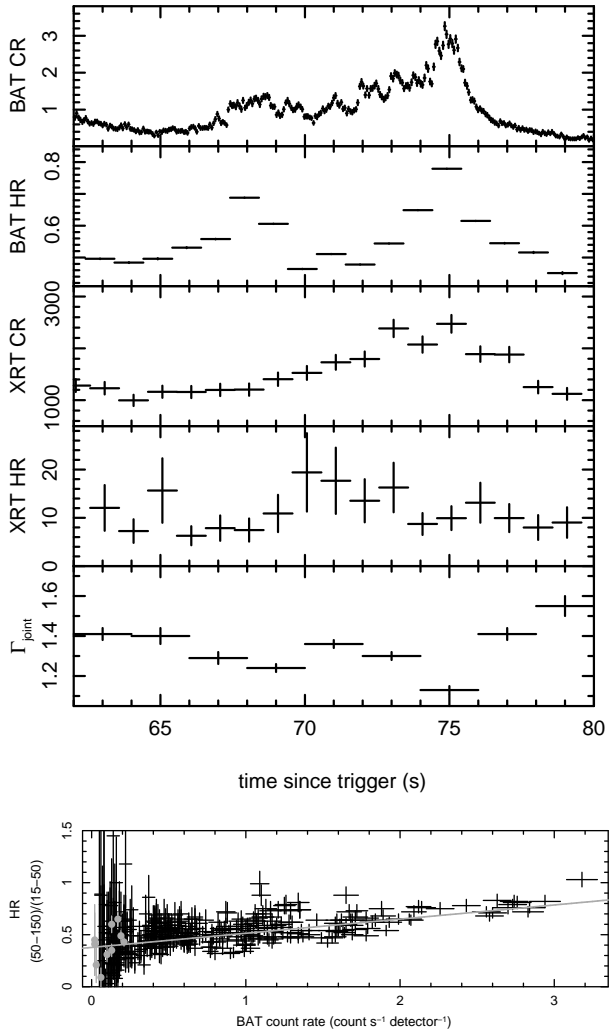

\begin{center}
\includegraphics[clip, width=8.0cm]{f2a.ps}

\includegraphics[clip, angle=-90,width=8.0cm]{f2b.ps}
\caption{Top panels: Light-curves, hardness ratios (HR) and the variation in $\Gamma$ using
  a single power-law fit during the main emission.  The BAT light-curve (top panel) is in
  units of count~s$^{-1}$~(fully illuminated detector)$^{-1}$, and the
  corresponding hardness ratio plots (50--150~keV)/(15--50~keV) using 1-s binning. The XRT
  light-curve shows counts over 0.3--10~keV, while the hardness ratio compares
  (1--10~keV)/(0.3--1~keV) over 1-s bins. Bottom panel: BAT hardness ratio
  versus count rate, showing that the emission is harder when brighter. Data
  from the precursor are shown as grey circles, with the main burst in
  black. The grey line shows a fit to the data, of the form HR~=~0.14~CR~+~0.39.}
\label{batxrthr}
\end{center}
\end{figure}


\paragraph{Spectral analysis}

For the precursor, T$_{\rm 90, pre}$~=~7.7~$\pm$~0.5~s
(15--150~keV). A spectrum extracted over this interval can be well modelled by a single power-law,
with $\Gamma$~=~1.68~$\pm$~0.09 ($\chi^{2}$/dof~=~26.2/23); no significant
improvement was found by using the Band
function (Band et al.\ 1993) or a cut-off power-law and a thermal model led to a
 slightly ($\chi^{2}$~$\sim$~8) worse fit. The 15--150~keV fluence for this time interval is
4~$\times$~10$^{-7}$ erg~cm$^{-2}$.

Considering only the main event,
T$_{\rm 90, main}$~=~18.2~$\pm$~1.1~s (measured from 61.8--80.0~s post-trigger). Fitting a power-law to the mean spectrum
during this time also results in a
good fit ($\Gamma$~=~1.40~$\pm$~0.01; fluence = 1.31~$\times$~10$^{-5}$
erg~cm$^{-2}$ over 15--150~keV; $\chi^{2}$/dof~=51.6/56 ); again, neither the
Band function nor a cut-off power-law improves upon this. There is significant
spectral evolution during the T$_{\rm 90}$ period, as shown in Figure~\ref{batxrthr}: at times when the count rate
is higher, the spectrum is harder. This behaviour was also common in earlier
bursts, as well as previous {\it Swift} detections (e.g. Golenetskii et al.\
1983; Ford et al.\ 1995; Borgonovo \& Ryde 2001; Goad et al.\ 2007). The precursor
shows a similar dependence of hardness ratio on count rate, suggesting that
the emission processes in the precursor and the main burst are the same or similar.

\subsubsection{{\it Konus-Wind}}
\label{sec:kw}

\paragraph{Temporal Analysis}
 {\it Konus-Wind} (Aptekar et al.\ 1995) triggered on the main episode of GRB~061121, while {\it Konus-A} triggered on the precursor (Golenetskii et al.\ 2006). Because of the spatial
separation of {\it Swift} and {\it Wind}, the light travel-time between
the spacecraft is 1.562~s: the {\it Konus-Wind} trigger
time, T$_{\rm 0,K-W}$~=~T$_{\rm 0,BAT}$~+~61.876~s. All {\it Konus} light-curves have
been plotted with respect to the BAT trigger, corrected for the light
travel-time. Figure~\ref{allbandlc} shows the {\it Konus-Wind} data plotted
over the standard energy bands, with 64~ms binning; the bottom
panel plots the coarser time resolution (2.944~s) `waiting mode' data, showing
that {\it Konus-Wind} did see slightly enhanced emission at the time of the precursor. The background levels (which have been subtracted in each case)
were 1005, 370 and 193.4~count~s$^{-1}$ for bands 21--83, 83--360 and
360--1360~keV, respectively.

\paragraph{Spectral analysis}
Table~\ref{kwtab} gives the spectral fits to the {\it Konus-Wind} data in
three separate time intervals shown by vertical lines in
Figure~\ref{allbandlc} ({\it Konus-Wind} spectral intervals are automatically
selected onboard): up to the end of the `bump' around 70~s (the `start' of the
burst), the burst maximum and, finally, until most of the emission has died
away (the burst tail). The data
were fitted with a cut-off power-law, where dN/dE~$\sim$~
E$^{-\Gamma}$~$\times$~e$^{[-(2-\Gamma)E/E_{peak}]}$, leading to the photon indices
and peak energies given in the table. The Band function was used to estimate upper limits for
the photon index above the peak; the values for the
peak energy and $\Gamma$ obtained from the Band function were the same as when
fitting the cut-off power-law. Little variation in the spectral slope for
energies below
the peak is seen over these intervals, though the peak itself may have moved to somewhat higher
energies during the burst emission. Extracting BAT spectra over the same time
intervals, and fitting with the same model (fixing E$_{\rm peak}$ at the value
determined from the {\it Konus-Wind} data) results in consistent
spectral indices.

\begin{table*}
\begin{center}
\begin{tabular}{lccccc}
\tableline
\tableline
start time (s) & stop time (s) &$\Gamma$ & E$_{peak}$ (keV) & $\Gamma_{\rm Band}$ & $\chi^{2}$/dof\\
\tableline
 61.876 & 70.324 & 1.40$^{+0.08}_{-0.09}$ & 478$^{+158}_{-99}$ &   $<$2.1 & 72/75\\
  70.324 & 75.188 & 1.23$^{+0.05}_{-0.06}$ &  608$^{+87}_{-71}$ &  $<$2.9 & 88/75\\
  75.188 & 83.380 & 1.30$^{+0.11}_{-0.13}$ &  621$^{+282}_{-159}$ &  $<$2.3 & 81/75\\
\tableline
61.876 &  83.380 & 1.32$^{+0.04}_{-0.05}$ &  606$^{+90}_{-72}$ &   $<$2.7 & 95/75\\
\tableline
\end{tabular}
\caption{{\it Konus-Wind} cut-off power-law spectral fit results. Times are given with respect to the BAT
  trigger. $\Gamma_{\rm Band}$ is the upper limit obtained for the spectral
  index above E$_{peak}$ when fitting with the Band function.}
\label{kwtab}
\end{center}
\end{table*}

\subsection{X-ray Data}

\subsubsection{XRT}
\label{sec:xrt}

\paragraph{Temporal Analysis}

The XRT identified and centroided on an uncatalogued X-ray source in a 2.5~s
Image Mode (IM) frame, as soon as the instrument was on target. This was
quickly followed by a pseudo Piled-up Photo Diode (PuPD) mode frame.
Following damage from a micrometeoroid impact in May 2005 (Abbey et al.\ 2005), the Photo Diode mode
(Low Rate and Piled-up) has been disabled [see Hill
et al.\ (2004) for details on the different XRT modes]; however, the XRT team
are currently working on a method to re-implement these science modes and to
update the ground software to process the files. The pseudo PuPD point
presented here is the first use of such data.

Data were then collected
in Windowed Timing (WT) mode starting at a count rate of
$\sim$~1280 count~s$^{-1}$ (pile-up
corrected -- see below); the rate rapidly increased to a maximum of
$\sim$~2500 count~s$^{-1}$ at T$_{0}$~+~75~s, making GRB~061121 the brightest
burst yet detected by the XRT. Following this peak, the
count-rate decreased, with a number of small flares superimposed on the
underlying decay (see Figure~\ref{allbandlc}). Photon Counting (PC) mode was automatically selected when the count
rate was below about
10~count~s$^{-1}$. Around 1.5~ks, the XRT switched back into WT mode
briefly, due to an enhanced background linked to the sunlit Earth and a
relatively high CCD temperature.

Because of the high count rate, the early WT data were heavily
piled-up; see Romano et al.\ (2006) for information about pile-up in this
mode. To account for this, an extraction region was used which excluded
the central 20 pixels (diameter; 1~pixel~=~2$\farcs$36) and extended out to a total
width of 60 pixels. Likewise, the first three
orbits of PC data were piled-up, and the data were thus extracted using
annular regions (inner exclusion diameter decreasing from 12 to 6 to 4 pixels as
the afterglow faded; outer diameter 60 pixels).  
The count rate was then corrected for the excluded
photons by a comparison of the Ancillary Response Files (ARFs) generated with
and without a correction for the Point Spread Function (PSF); the ratio of these
files provides an estimate of the correction factor. Nousek et al.\ (2006)
give more details on this method. Occasionally, the afterglow was partially
positioned over the CCD columns disabled by micrometeoroid damage mentioned above. In these cases, the data were corrected using an exposure map.

From T$_{0}$~+~3~$\times$~10$^{5}$~s onwards, the afterglow had faded
sufficiently for a nearby (41$\farcs$5 away), constant (count rate~ $\sim$~0.003 count~s$^{-1}$) source to contaminate the
GRB region; this source is coincident with a faint object in the Digitized Sky
Survey and is
marginally detected in the UVOT $V$ filter. Thus, beyond this time, the extraction region was decreased
to a diameter of 30~pixels, and the count rates corrected for the loss in PSF (a
factor of $\sim$~1.08). The spectrum of this nearby source can be
modelled with a single power-law of $\Gamma$~=~1.5$^{+0.2}_{-0.1}$, with
N$_{\rm H}$~=~(1.8$^{+1.6}_{-1.2}$)~$\times$~10$^{21}$~cm$^{-2}$, in comparison with the Galactic value in this direction of
5.09~$\times$~10$^{20}$~cm$^{-2}$ (Dickey \& Lockman 1990).

\begin{figure}
\begin{center}
\includegraphics[clip, angle=-90,width=8.0cm]{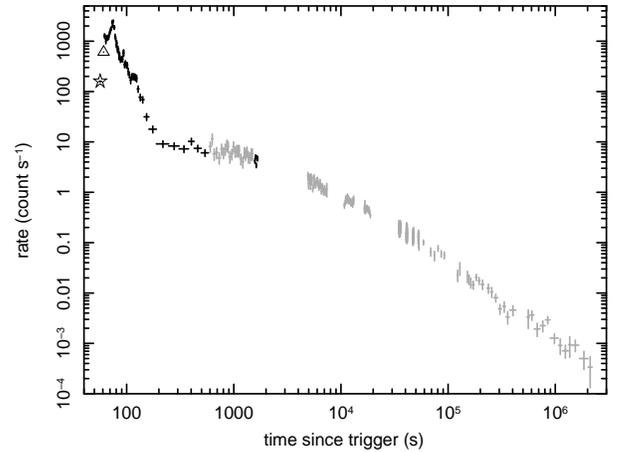}
\caption{{\it Swift}-XRT light-curve of GRB~061121. The
  star and triangle show the initial Image Mode and pseudo PuPD point
  (see text for details), followed by WT mode data (black) during the main
  burst (and at the end of the first orbit) and PC mode data (in grey).}
\label{wtimlc}
\end{center}
\end{figure}

Figure~\ref{wtimlc} shows the XRT light-curve, starting with the IM point (see Hill et al.\ 2006 for details on how IM data
are converted to a count rate) and followed by the pseudo PuPD mode data. The importance of
these early pre-WT data is clear, confirming that the XRT caught
the rise of the main burst. 


\begin{figure}
\begin{center}
\includegraphics[clip, angle=-90,width=8.0cm]{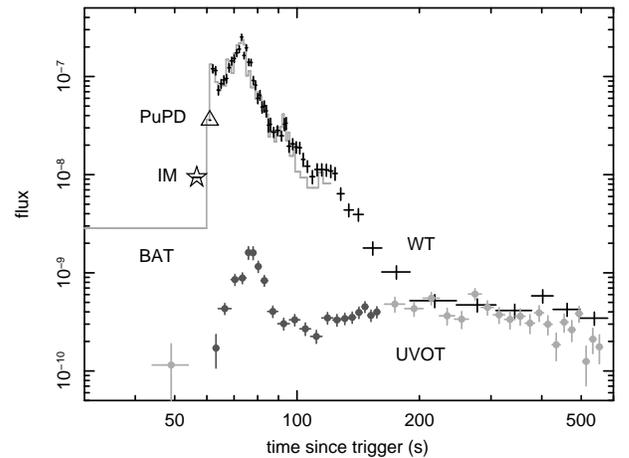}
\caption{{\it Swift} flux light-curve of GRB~061121, showing the early X-ray
  data (star, triangle and crosses) and the BAT data (grey histogram) extrapolated into the 0.3--10~keV
  band pass in units of erg~cm$^{-2}$~s$^{-1}$, together with the UVOT
  flux density light-curve (light grey circles -- $V$-band; dark grey circles --
  White filter) in units of
  erg~cm$^{-2}$~s$^{-1}$~$\rm \AA^{-1}$, scaled to match the XRT flux observed
  at the start of the
  `plateau' phase.}
\label{joint}
\end{center}
\end{figure}


After the bright burst, the afterglow began to follow the
`canonical' decay, seen in many {\it Swift} bursts (Nousek et
al. 2006; Zhang et al.\ 2006a). Such a decay
can be parameterised by a series of power-law segments;
in this case, fitting the data beyond 200~s after the trigger (=~125~s after
the main peak), two breaks in
the light-curve were identified, with the decay starting
off very flat ($\alpha$~=~0.38~$\pm$~0.08) and eventually steepening
to $\alpha$~=~1.07$^{+0.04}_{-0.06}$ at $\sim$~2.3~ks and then
$\alpha$~=1.53$^{+0.09}_{-0.04}$ at $\sim$~32~ks
(Table~\ref{xrtlc}). The addition of the second break vastly
improved the fit by $\Delta\chi^2$~=~112.4 for two degrees of freedom.
However, we note that O'Brien et al.\ (2006) and Willingale et al.\ (2007)
advocate a different description of the temporal decline; we return to this in
$\S$\ref{disc}.

Fitting the decay of the main peak (75--200~s, keeping T$_{0}$ as the trigger
time) with a power-law, the slope is very steep, with
$\alpha_{0}$~=~5.1~$\pm$~0.2. However, both Zhang et al.\ (2006a) and Liang et
al.\ (2006) have shown
that the appropriate time origin is the start
of the last pulse. Thus, a model of the form
f(t)~$\propto$~(t$-$t$_{0}$)$^{-\alpha_{0}}$ was used, finding t$_{0}$~=~58~$\pm$~1~s and a slope of $\alpha_{0}$~=~2.2$^{+0.4}_{-0.3}$; this is a statistically significant
improvement on the power-law fit using the precursor T$_{0}$
($\Delta\chi^{2}$~=~32 for one extra parameter).

\begin{table}
\begin{center}
\begin{tabular}{lcc}
\tableline
\tableline
$\alpha_{1}$  & 0.38~$\pm$~0.08 & Plateau phase\\
T$_{\rm break, 1}$ & 2258$^{+507}_{-377}$~s\\
$\alpha_{2}$ & 1.07$^{+0.04}_{-0.06}$ & Shallow phase\\
T$_{\rm break, 2}$ & (3.2$^{+2.1}_{-0.6}$)~$\times$~10$^{4}$~s\\
$\alpha_{3}$  & 1.53$^{+0.09}_{-0.04}$ & Steep phase\\
\tableline
\end{tabular}
\caption{XRT power-law light-curve fits from 200~s after the trigger
  onwards; times are referenced to the BAT trigger. The names used in the text for the different epochs of the light-curve are listed in the last column.}
\label{xrtlc}
\end{center}
\end{table}

Figure~\ref{joint} plots the {\it Swift} data in terms of flux
(the BAT data have been extrapolated into the 0.3--10~keV band, using the
joint fits with the XRT described in $\S$\ref{sec:bat-xrt}) and flux density for UVOT. The BAT and XRT
data are fully consistent with each other at all overlapping times.

 \paragraph{Spectral Analysis}

\begin{figure}
\begin{center}
\includegraphics[clip, angle=-90,width=8.0cm]{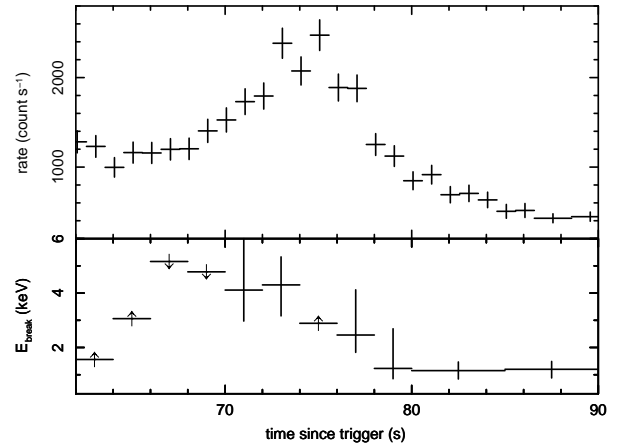}
\caption{Fitting the X-ray data over 0.3--10~keV with a broken power-law
  ($\Gamma_{1}$~=0.69$^{+0.13}_{-0.07}$ and $\Gamma_{2}$~=1.61$^{+0.14}_{-0.13}$  for all spectra), the break energy
  seems to move through the band, towards higher energies when the emission is
  brighter. Arrows indicate upper or lower 90\% limits.}
\label{xrtbkn}
\end{center}
\end{figure}

The XRT data also show that strong spectral evolution was present throughout the period of the prompt
emission; this is discussed in conjunction with the BAT data in
$\S$\ref{sec:bat-xrt}. Considering the X-ray data alone, there is some
indication that the spectra may be better modelled with a broken, rather than single, power-law, although the break energies cannot always be well
constrained (see Figure~\ref{xrtbkn}). For each spectrum [covering periods of 2~s during
the main pulse, followed by two spectra of 5~s (80--85~s and 85--90~s) where
the emission is fainter], the low-energy slopes were tied together for each
spectrum (i.e., the
slope measured is that averaged over all of the spectra), as were the high-energy indices, and the rest-frame column density, N$_{\rm H,z}$, was fixed at
(9.2~$\pm$~1.2)~$\times$~10$^{21}$ cm$^{-2}$ from the best fit to the
data from later times (see below); only the break energy and the
normalisation were allowed to vary. When simultaneously fitting all 11
spectra, $\chi^{2}$/dof decreased from 142/134 to 127/132. Individually, the
spectral fits were typically improved by $\chi^2$ of between 2--5.


The X-ray data during the GRB~051117A
flares (Goad et al.\ 2007) were found to be better modelled with broken
power-laws, with the break energy moving to harder energies during each flare
rise, and then softening again as the flux decayed. Likewise, Guetta et
al.\ (2006) found breaks in the X-ray spectra obtained during the flares in
GRB~050713A. The same pattern may be occurring here, and there is certainly an
indication of spectral curvature. 

The observed flux calculated from the spectrum corresponding to
the peak of the emission (74--76~s) was measured to be 1.66~$\times$~10$^{-7}$ erg~cm$^{-2}$~s$^{-1}$ (over 0.3--10~keV); the
unabsorbed value was 1.77~$\times$~10$^{-7}$ erg~cm$^{-2}$~s$^{-1}$.


The PC spectra
were also extracted for the various phases of the light-curve (`plateau',
`shallow' and `steep' -- defined in Table~\ref{xrtlc}); the results of the
fitting are
presented in Table~\ref{xrtspec}. In each phase, the spectrum could be well modelled by
a single power-law (no break required), with excess absorption in the rest-frame of the GRB (modelled using {\sc ztbabs} and the `Wilms' abundance in {\sc
  xspec}; Wilms et al.\ 2000).
Together with the WT spectrum from
$\sim$~200--590~s after the trigger (in the plateau stage), the first two PC
spectra (plateau and shallow) are fully consistent with a
constant photon index of $\Gamma$~=~2.07~$\pm$~0.06 and
N$_{\rm H,z}$~=~(9.2~$\pm$~1.2)~$\times$~10$^{21}$ cm$^{-2}$. 

Following the
second apparent break in the light-curve, around 3.2~$\times$~10$^{4}$~s, the spectrum hardened slightly, to
a photon index of $\Gamma$~=~1.83~$\pm$~0.11 (or 1.87~$\pm$~0.08 using
N$_{\rm H,z}$~=~9.2~$\times$~10$^{21}$ cm$^{-2}$).

\begin{table*}
\begin{center}
\begin{tabular}{lccccc}
\tableline
\tableline
Epoch & time since  & $\Gamma$ & N$_{\rm H,z}$ & $\chi^{2}$/$\nu$ &
corresponding \\
& trigger (s) & & (10$^{21}$ cm$^{-2}$) & & $\alpha$\\
\tableline

Plateau & 590--1560 & 2.14~$\pm$~0.12 &  10.8$^{+2.5}_{-2.8}$ & 62.5/52 & 0.38~$\pm$~0.08\\
Shallow & 4900--22245 & 2.04~$\pm$~0.10 & 8.9$^{+2.1}_{-2.4}$ & 67.5/70 & 1.07$^{+0.04}_{-0.06}$\\
Steep & 34550--1152750 & 1.83~$\pm$~0.11 & 8.0$^{+2.6}_{-2.2}$ &
48.0/55 & 1.53$^{+0.09}_{-0.04}$\\
\hline
Plateau & 590--1560 & 2.09~$\pm$~0.08 &  9.2~$\pm$~1.2 (tied) & 63.5/53 & 0.38~$\pm$~0.08\\
Shallow & 4900--22245 & 2.05~$\pm$~0.06 & 9.2 (tied) & 67.6/71 & 1.07$^{+0.04}_{-0.06}$\\
Steep & 34550--1152750  & 1.87~$\pm$~0.08 & 9.2 (tied) & 48.7/56 &
1.53$^{+0.09}_{-0.04}$\\
\tableline
\end{tabular}
\caption{XRT PC spectral fits - rest-frame N$_{\rm H}$ free and then tied between
  all three spectra. The temporal decay slopes, $\alpha$, corresponding to
  each stage are also given. The Galactic absorbing column of
  N$_{\rm H}$~=~5.09~$\times~$10$^{20}$ cm$^{-2}$ was always included in the model.}
\label{xrtspec}
\end{center}
\end{table*}

\subsubsection{{\it XMM-Newton}}
\label{xmm}

{\it XMM-Newton} (Jansen et al.\ 2001) performed a Target of Opportunity
observation of GRB~061121 (Observation ID 0311792101)
less than 6.5~hr after the trigger (Schartel 2006) and collected data for
$\sim$~38~ks (MOS1, MOS2; Turner et al.\ 2001) and $\sim$~35~ks (PN; Str{\"
  uder} et al.\ 2001). This observation
is mainly during the `shallow' phase, though also covers a short timespan
after the break at around 32~ks. 

\begin{figure}
\begin{center}
\includegraphics[clip, angle=-90,width=8.0cm]{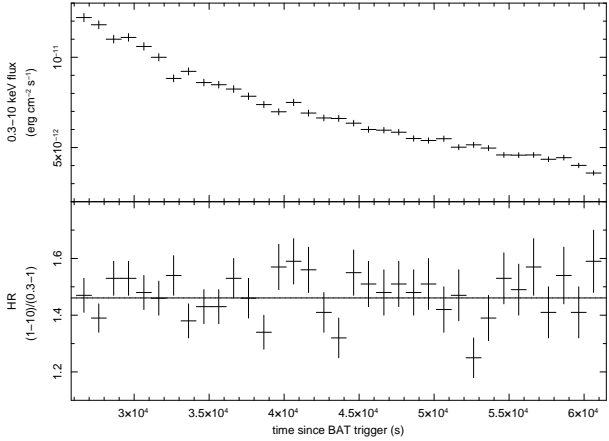}
\caption{{\it XMM-Newton} EPIC-PN light-curve and hardness ratio of
  GRB~061121. The horizontal line shows the hardness ratio is consistent with
  a constant value of $\sim$~1.46, indicating there is no spectral
  evolution during this time. }
\label{pnlc}
\end{center}
\end{figure}

Figure~\ref{pnlc} plots the PN flux light-curve and hardness ratio during the {\it
  XMM-Newton} observation, showing the lack of spectral evolution during this
  time frame; a hardness ratio calculated for the {\it Swift} data was in
  agreement with this finding. The decay slope over
  this time (MOS1, MOS2, PN and joint) is consistent with the {\it Swift}
  results ($\alpha$~$\sim$~1.3; note this crosses the time of the second break
  in the decay).

\begin{figure}
\begin{center}
\includegraphics[clip, angle=0,width=9.0cm]{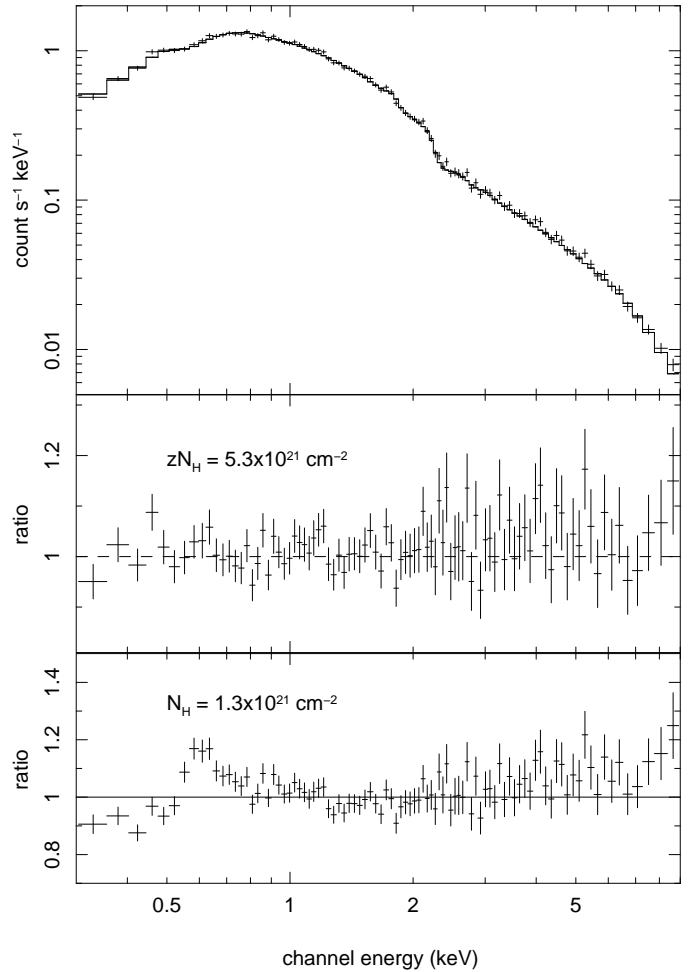}
\caption{EPIC-PN spectrum of the late-time afterglow of GRB~061121, with an
  excess absorbing column both in the rest-frame of the GRB and the observer's
  frame. The spectrum is much better modelled with an excess column at z~=~1.314.}
\label{nh}
\end{center}
\end{figure}

The {\it XMM-Newton} EPIC\footnote{European Photon Imaging Camera} spectra show clear
evidence for excess N$_{\rm H}$, in agreement with the {\it Swift}
data. In addition, fitting with excess N$_{\rm H}$ in the rest-frame
of the GRB gives a significantly better fit than at z~=~0, as shown in
Figure~\ref{nh}. When fitting in
the observer's frame there is a noticeable bump in the residuals around
0.6~keV; fitting with N$_{\rm H}$ at z~=~1.314 removes this feature. The data
are of sufficiently high signal-to-noise that the 
redshift of the absorber can be estimated from the spectrum. Limits can be placed on the redshift and
absorbing column, respectively, of z~$>$~1.2 and N$_{\rm
  H,z}$~$>$~4.6~$\times$~10$^{21}$~cm$^{-2}$ at 99\% confidence, in
agreement with the spectroscopic redshift from Bloom et al.\ (2006) within the
statistial uncertainties. At
their value of z~=~1.314, the excess N$_{\rm H,z}$ from the
EPIC-PN spectrum
is (5.3~$\pm$~0.2)$\times$~10$^{21}$~cm$^{-2}$, lower than
the best fit to the {\it Swift} data from the
simultaneous `shallow' decay section, but more similar to the values
obtained from fitting the optical-to-X-ray Spectral Energy Distributions
(SEDs) in $\S$\ref{sec:uvot-xrt}. In agreement with the simultaneous
XRT PC mode data, there is no evidence for a break in the EPIC spectrum over
this time period. Spectra from neither the Reflection Grating Spectrometer
(den Herder et al.\ 2001) nor EPIC show obvious absorption or emission lines.


\subsubsection{{\it Chandra}}
\label{chandra}

Chandra performed a 33~ks Target of Opportunity observation at $\sim$~61 day after the trigger. No
source was detected at the position of the X-ray afterglow, with a 3$\sigma$ upper
limit of 2.5~$\times$~10$^{-15}$~erg~cm$^{-2}$~s$^{-1}$.

\subsection{Optical/UV Data}

\subsubsection{UVOT}

The UVOT detected an optical counterpart in the initial White
filter\footnote{The White filter covers a broad bandpass of $\lambda\sim 1600-6500$~\AA.}
observation, starting 62~s after the trigger, and subsequently in all other
filters (optical and UV). The UVOT followed the typical sequence for GRB
observations, with the early data being collected in event mode, which has
a frame time of 8.3~ms during this observation.\footnote{The data have been
  adjusted to take into account an
incorrect onboard setting (between 2006-11-10 and 2006-11-22), which resulted in the wrong frame times being stored
in the headers of the UVOT files (Marshall 2006a).}
Photometric measurements were obtained from the UVOT
data using a circular source extraction region with a $5-6\arcsec$ radius. {\sc
  uvotmaghist} was used to convert count rates to magnitudes and flux; no
normalisation between the different filters was applied.

As in the gamma-ray and X-ray
bands, the main burst was detected, with an increase in count rate seen
between $\sim$~50--75~s after the trigger (Figures~\ref{allbandlc} and \ref{joint}). However, although an increase in count rate is seen for the UVOT data, it is
by a smaller factor than observed for the XRT. After $\sim$~110~s, the UVOT emission stops decaying and rebrightens slightly,
until 140~s after the trigger, at which time it flattens off and then starts
to fade again (Figure~\ref{joint}).  The slower decay between
$\sim$~100--200~s may be indicative of the contribution of an additional
(afterglow) component beginning to dominate.

A single UV/optical light curve was created from all the UVOT filters in order
to get the best measurement of the optical temporal decay. This was done by
fitting each filter dataset individually (between 200 and 1~$\times$~10$^{5}$~s) and finding the normalisation, which was then
modified to correspond to that of the $V$-band light-curve.
The decay across all the filters beyond 200~s after the trigger can be fitted with a single
slope of $\alpha_{\rm UVOT}$~=~0.68~$\pm$~0.02; the individual $U$, $B$ and $V$ decay
rates are consistent with one another. No break in the light-curve is
seen out to $\sim$~100~ks.


\subsubsection{{\it ROTSE}}
\label{rotse}

{\it ROTSE-IIIa}, at the Siding Spring Observatory in Australia, first imaged
GRB~061121 21.6~s after the trigger time under poor (windy) seeing conditions. A variable source
was immediately identified, at a position coincident with that determined by
the UVOT (Yost et al.\ 2006).

The {\it ROTSE} data (unfiltered, but calibrated to the $R$-band) have been
included in Figure~\ref{all-lc} (discussed later). It is noticeable that the
peak around 75~s seen in the {\it Swift} data is not readily apparent in the
{\it ROTSE} measurements. The bandpass of the UVOT White filter is
more sensitive to photons with wavelengths of
$\lambda$~$<$~4500\AA\footnote{See
  http://swiftsc.gsfc.nasa.gov/docs/heasarc/caldb/swift/}, while the {\it ROTSE}
bandpass is redder. This, together with poor seeing conditions during the observation, may explain
why the {\it ROTSE} light-curve does not clearly show the main emission.

\subsubsection{Faulkes Telescope North}
\label{ftn}

The FTN, at Haleakala on Maui, Hawaii, began observations of
GRB~061121 225~s after the burst trigger, performing a $BVRi'$ multi-colour
sequence (Melandri et al.\ 2006). $R$-band photometry was
performed relative to the USNO-B 1.0 `R2' magnitudes. Magnitudes were
then corrected for Galactic  extinction using the dust-extinction maps
by Schlegel et al.\ (1998),  and converted to fluxes using the absolute
flux calibration from Fukugita et al.\ (1995).
The photometric $R$-band points have been included in Figure~\ref{all-lc}.

\subsection{Broadband Modelling}
\label{sec:broad}

\subsubsection{Gamma-rays -- X-rays}
\label{sec:bat-xrt}

\paragraph{Spectral Analysis}

Because the BAT was in event mode throughout the observation of the main burst
of GRB~061121, detailed spectroscopy could be performed. Unfortunately this was not the case during the prompt observation of GRB~060124 (Romano et al.\ 2006).

Figure~\ref{batxrthr} demonstrates the spectral evolution seen
in both the BAT and XRT during the prompt emission. Spectra were extracted over 2~s intervals, in an attempt to obtain
sufficient signal to noise while not binning over too much of the rapid
variability. The BAT data are hardest around 68~s and 75~s (the second of
these times corresponding to the peak of the main emission); the XRT hardness
peaks about 70~s, which could be a further indication of the softer data
lagging the harder. The joint spectrum
($\Gamma_{\rm joint}$ comes from a simple absorbed power-law fit to the simultaneous BAT and XRT data)
is at its
hardest during the brightest part of the emission. The joint fit also hardens
around 68--70~s, between the times when the BAT and XRT data respectively are
at their hardest. The onboard spectral time-bin selection prevents the {\it Konus-Wind} data from being sliced into corresponding times, so
constraints have not been placed on the high energy cut-off, E$_{\rm
  peak}$. Breaks in the XRT-BAT power-laws can only be poorly constrained.

In Figure~\ref{joint}, the BAT and XRT data were converted to 0.3--10~keV
fluxes using the time-sliced power-law fits to the simultaneous BAT and XRT
spectra. Without the use of such varying conversion factors, the derived BAT
and XRT fluxes would have been inconsistent with each other.

A broadband spectrum, covering 0.3~keV to 4~MeV in the observer's frame (XRT, BAT and {\it Konus-Wind}) for
$\sim$~70--75~s post trigger was fitted by the absorbed cut-off power-law model
described in $\S$\ref{sec:kw}. A constant factor of up to 10\% was included
between the BAT and {\it Konus-Wind} data, to allow for calibration
uncertainties. The best fit ($\chi^2$/dof~=~301/167) gives $\Gamma$~=~1.19~$\pm$~0.01, with E$_{\rm
  peak}$~=~670$^{+65}_{-47}$~keV. N$_{\rm H,z}$ was fixed at
9.2~$\times$~10$^{21}$~cm$^{-2}$ (from the X-ray fits in
$\S$\ref{sec:xrt}). 
Allowing $\Gamma$ to vary between the three spectra hints
at further spectral curvature, although the differences are
marginal, significant at only the 2$\sigma$ level.

The isotropic equivalent energy (calculated using the time-integrated flux
over the full T$_{90}$ period) is 2.8~$\times$~10$^{53}$~erg in the 1~keV --
10~MeV band (GRB rest frame), meaning that GRB~061121 is consistent with the
Amati relationship (Amati et al.\ 2002). See $\S$\ref{rwmodel} for a
beaming-corrected gamma-ray energy limit.

\paragraph{Lag Analysis}
\label{lag}

A lag analysis (e.g., Norris et al.\ 1996) between the BAT bands leads to interesting
results. 
Comparing bands 50--100~keV and 15--25~keV, the precursor emission 
yields a spectral lag of 600~$\pm$~100~ms, while the main emission has a much
smaller lag of 1~$\pm$~6~ms. Note that the calculation was performed using 64~ms binning for the precursor
and 4~ms binning for the main burst; see Norris (2002) and Norris \& Bonnell (2006) for
  more details on the procedure.
This lag for the main emission is rather small
for a typical long burst, however both lags are consistent with the long-burst luminosity-lag
relationship generally seen (Norris et al.\ 2000). The
short spectral lag for the main emission, and the longer value for the
precursor are also found when comparing the 100--350~keV and 25--50~keV
bands.

Similarly, comparison of the hard and soft (2--10~keV and 0.3--2~keV) XRT
bands reveals a lag of approximately 2.5~s, as the emission softens through
the main burst. The X-ray data also lag behind the gamma-ray data, and the
optical behind the X-ray.

\begin{figure}
\begin{center}
\includegraphics[clip, angle=-90,width=8.0cm]{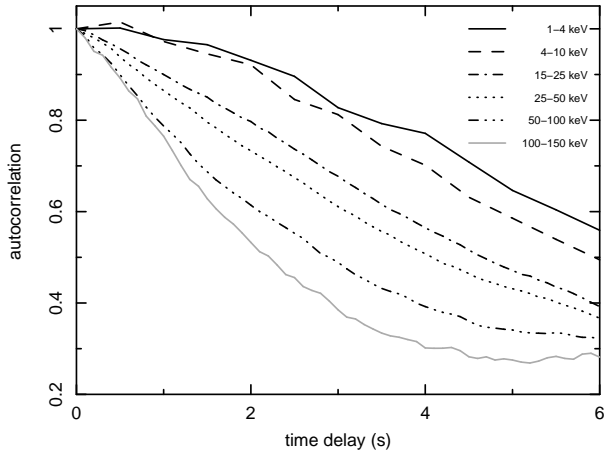}
\caption{Autocorrelation function of the BAT and XRT data during the prompt
  emission of GRB~061121, showing that the main burst peak is
  broader at softer energies.}
\label{autocorr}
\end{center}
\end{figure}

Link et al.\
(1993) and Fenimore et al.\ (1995) used a sample of BATSE\footnote{Burst And
  Transient Source Experiment} (Paciesas et al.\ 1999) bursts to investigate
the relationship between the duration of bursts and the energy band
considered.  They found that the bursts, and smaller structures within the main emission, generally
become shorter with increasing energy (see also Cheng et al.\ 1995; Norris et al.\ 1996; in't Zand
\& Fenimore 1996; Piro et al.\ 1998). Figure~\ref{autocorr} plots the autocorrelation function over various
X-ray and gamma-ray bands, to reinforce the point that the peak is narrower
the harder the band -- over X-ray as well as gamma-ray energies. Comparison of
the light-curves over the different energy bands in Figure~\ref{allbandlc}
demonstrates this as well. A similar behaviour was also found for GRB~060124,
where Romano et al.\ (2006) compared the T$_{90}$ values obtained
for the main burst over the X-ray and gamma-ray bands. Fenimore et al.\ (1995)
found that the width of the autocorrelation function, W~$\propto$~E$^{-0.4}$,
where E is the energy at which the function was determined; the six measurements from
GRB~061121 are consistent with this finding. 


\subsubsection{Optical -- X-rays}
\label{sec:uvot-xrt}

Using the {\it Swift} X-ray and UV/optical data, $R$ and $i'$ band data from the Faulkes
Telescope and $Rc$ data from the Kanata telescope (Uemera et al.\ 2006), SEDs were
produced at epochs corresponding to the peak of the emission (72--75~s post BAT trigger), the
plateau stage and during the shallow decay. Fitting at the different epochs gives an estimation of the broadband spectral
variation.

 For each of the UVOT lenticular filters, the tool {\sc
  uvot2pha} was used to produce spectral files compatible with {\sc
  xspec}, and for the latter two epochs the count rate in each band
was set to that determined from a power-law fit to the individual filter light
curves over the time interval in question, using $\alpha$~=~0.68. To determine the Faulkes Telescope $R$ and $i'$ band flux
during the plateau stage, a power law was fitted to the complete data set
(220--1229~s post BAT trigger for $R$ and 467--1401~s for $i'$) with the decay
index left as a free parameter. The $R$ magnitude at the mid-time of the
shallow stage (6058~s) was determined from the Kanata $R$-band magnitude
reported at 6797~s (Uemera et al.\ 2006), assuming the same decay index as observed in the
UVOT data.  An uncertainty of 0.2~mag was assumed as the
systematic uncertainty for the photometric calibration of the ground based
data.
 
At a redshift of z~=~1.314, the beginning of the Lyman-$\alpha$ forest is
redshifted to an observer-frame wavelength of $\sim 2812$~\AA\, which falls
within the UVW1 filter bandpass, the reddest of the UV filters. A correction
was applied to the three UV filter fluxes to account for this
absorption, based on parameters from Madau (1995) and Madau et al.\ (1996); see
also Curran et al.\ (in prep). 

The methods used for simultaneous fitting of the SED components are
described in detail in Schady et al.\ (2007a).
The SEDs were fitted with a power-law, or a broken power-law, as expected from the
synchrotron emission, and two dust and gas
components, to model the Galactic and host galaxy photoelectric absorption and
dust extinction. The column density and reddening in the first absorption
system were fixed at the Galactic values. [The Galactic extinction along this
  line of sight is $E(B-V)=0.046$ (Schlegel et al.\ 1998).]

The second photoelectric absorption
system was set to the redshift of the GRB, and the neutral hydrogen column
density in the host galaxy was determined assuming Solar abundances. The
dependence of dust extinction on wavelength in the GRB host galaxy was
modelled using three extinction laws, taken from observations of the Milky Way
(MW), the Large Magellanic Cloud (LMC) and the Small Magellanic Cloud
(SMC) and parameterised by Pei (1992) and Cardelli et al.\ (1989). The greatest differences observed in these extinction laws are the
amount of far UV extinction (which is greatest in the SMC and least in
the MW) and the strength of the $2175$~\AA\ absorption feature (which is most
prominent in the MW and negligible in the SMC).

\begin{figure}
\begin{center}
\includegraphics[clip,width=8.5cm]{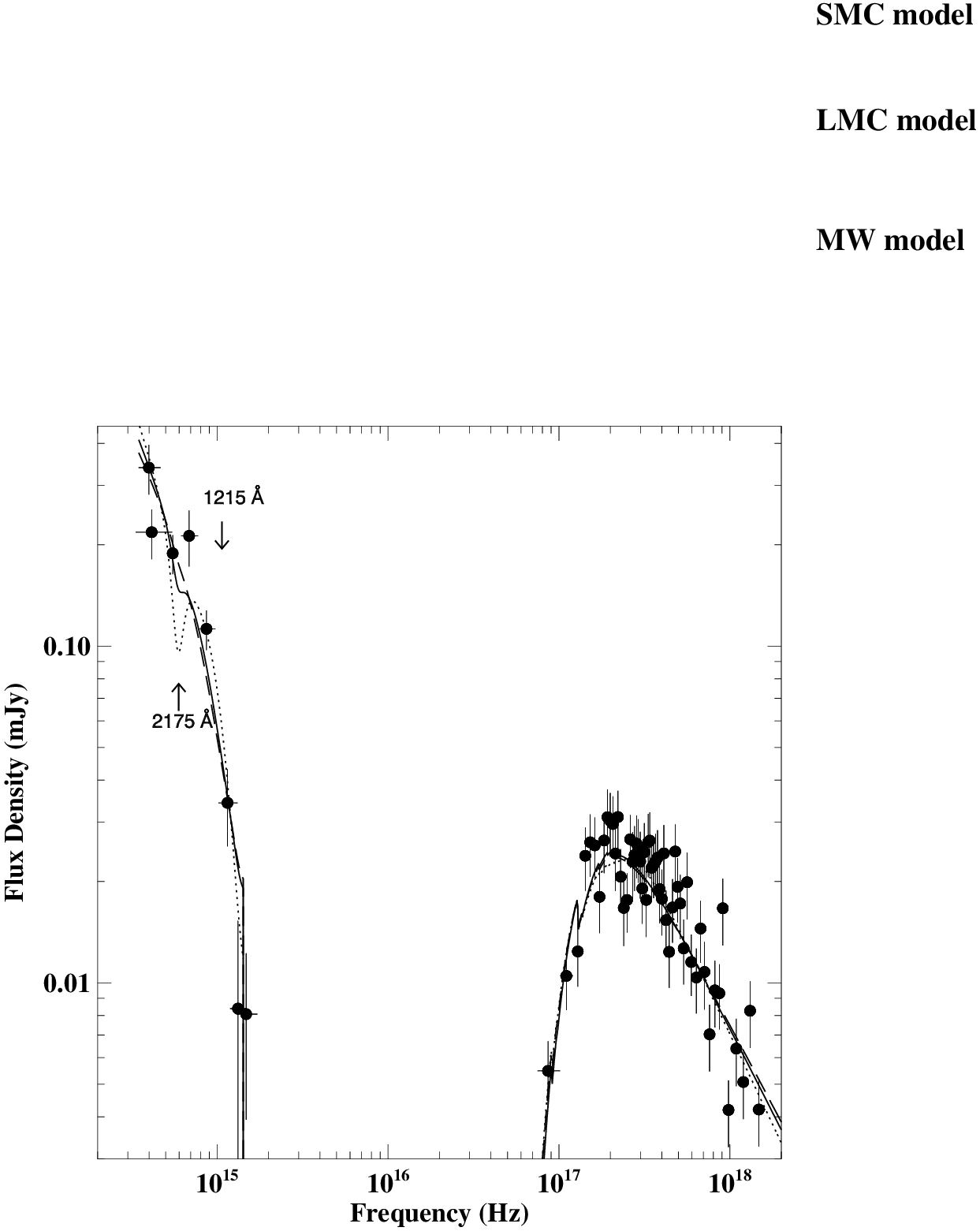}
\caption{Broken power-law fit to the UVOT, XRT and ground-based $R$ and $i'$
  spectral energy distribution of GRB~061121 between $\sim$~596--1566~s after the
  trigger (plateau phase) plotted in the observer's frame. The arrows indicate the beginning of the Lyman-$\alpha$ forest
  (1215\AA\ in the rest-frame) and the absorption feature in the MW dust
  extinction law (2175\AA), which is shown by a dotted line. The solid line corresponds to the
  LMC extinction, and the dashed one to the SMC extinction.}
\label{sedfit}
\end{center}
\end{figure}

Fitting these data together, a measurement of the spectral
slope and optical and X-ray intrinsic extinctions (for the second two epochs) were obtained
(Table~\ref{uvotsed}); the A$_{\rm V}$ values given in the table are in
addition to the A$_{\rm V}$~=~0.151 associated with the Milky Way itself. The slope above the break energy
(which lies towards the low energy end of the X-ray bandpass for each phase) was
assumed to be exactly 0.5 steeper than the spectral slope below the break
(the condition required for a cooling break), since allowing all of the
parameters to vary leads to unconstrained fits.
Figure~\ref{sedfit} shows, as an example, the fit to the data in the
plateau stage. 

A Milky Way dust extinction law provides the best overall fit
to the data, using a broken power-law model, although the LMC model is equally
acceptable.  

During the plateau phase, and adopting the broken power-law model parameters
given in Table~\ref{uvotsed}, we find
gas-to-dust ratios of (1.6~$\pm$~0.7), (2.6~$\pm$~0.7) and (3.0~$\pm$~0.7)
$\times$10$^{22}$~cm$^{-2}$~mag$^{-1}$ for MW, LMC and SMC fits respectively.
We can compare these estimates 
to the measured values for the MW of (4.93~$\pm$~0.45)~$\times$~10$^{21}$
cm$^{-2}$~mag$^{-1}$ (Diplas \&
Savage 1994) and the LMC and SMC of (2.0~$\pm$~0.8) and
(4.4~$\pm$~1.1)~$\times$10$^{22}$ cm$^{-2}$~mag$^{-1}$ , respectively
(Koornneef 1982; Bouchet et al.\ 1985). The MW fit to GRB~061121, which is
found to be marginally the best model, is consistent with the LMC gas-to-dust ratio only, at the 90\% confidence level. The ratios derived from the LMC and SMC fits are consistent with both the
LMC and SMC gas-to-dust ratios. We note that all fits are inconsistent
with the MW ratio at this confidence level, following the trend
seen in pre-{\it Swift} bursts (e.g., Starling et al.\ 2007 and references therein),
and that if a metallicity below Solar were adopted, the gas-to-dust
ratio of GRB~061121 would increase, moving it further towards the SMC
value.


\begin{table*}
\begin{center}
\begin{tabular}{lcccccccc}
\tableline
\tableline
X-ray & Model & Extinction &   N$_{\rm H,z}$ & $\Gamma_{1}$ & E$_{\rm break}$
& $\Gamma_{2}^{\ a}$ & A$_{\rm V} ^{\ b}$ & $\chi^2$/dof\\ 
Epoch & & & (10$^{21}$~cm$^{-2}$) & & (keV)\\
\tableline

Peak & PL & SMC & 1.6$^{+5.0}_{-1.6}$ & 0.99~$\pm$~0.01 & \nodata & \nodata & 0.64 &
25/27\\
& & LMC & 1.9$^{+5.3}_{-1.9}$ & 1.06~$\pm$~0.01 & \nodata & \nodata & 0.98 & 23/27\\
 & & MW & 2.4$^{+5.8}_{-2.4}$ & 1.16~$\pm$~0.01 & \nodata & \nodata & 1.51 & 22/27\\
\tableline
 & BKN PL & SMC & 2.7$^{+9.5}_{-2.7}$ & 0.72$^{+0.08}_{-0.15}$ &
0.17$^{+0.79}_{-0.15}$ & 1.22 & 0.51 & 22/26\\
& & LMC & 3.0$^{+7.7}_{-3.0}$ & 0.77$^{+0.08}_{-0.20}$ &
0.18$^{+0.53}_{-0.17}$ & 1.27 & 0.72 & 22/26\\
& & MW & 3.0$^{+7.7}_{-3.0}$ & 0.77$^{+0.10}_{-0.21}$ & 0.09$^{+0.30}_{-0.09}$
& 1.27 & 1.03 & 22/26\\
\tableline
\tableline
Plateau & PL & SMC & $1.42\pm 0.51$ & $1.58\pm 0.02$ & \nodata & \nodata & $0.62\pm 0.05$ & 167/59\\
 & & LMC & $1.98\pm 0.54$ & $1.64\pm 0.03$ & \nodata & \nodata & $0.94\pm 0.08$ & 152/59\\
 & & MW & $2.71\pm 0.69$ & $1.71\pm 0.03$ & \nodata & \nodata & $1.39\pm 0.10$ & 136/59\\
\tableline
 & BKN PL & SMC & $3.89^{+0.72}_{-1.01}$ & $1.46^{+0.03}_{-0.02}$ & $0.84^{+0.36}_{-0.12}$ & 1.96 & $0.52\pm 0.04$ & 84/58\\
 & & LMC & $4.40^{+0.77}_{-1.30}$ & $1.51^{+0.04}_{-0.02}$ & $0.82^{+0.16}_{-0.14}$ & 2.01 & $0.74\pm 0.06$ & 80/58\\
 & & MW & $3.91^{+0.77}_{-0.75}$ & $1.58^{+0.04}_{-0.03}$ & $1.22^{+0.25}_{-0.20}$ & 2.08 & $1.03^{+0.09}_{-0.08}$ & 79/58\\
\tableline
\tableline
Shallow & PL & SMC & $2.72\pm 0.49$ & $1.69\pm 0.02$ & \nodata & \nodata & $0.65\pm 0.04$ & 162/77\\
 & & LMC & $3.37^{+0.53}_{-0.49}$ & $1.75\pm 0.03$ & \nodata & \nodata & $0.98^{+0.07}_{-0.06}$ & 146/77\\
 & & MW & $4.60^{+0.65}_{-0.60}$ & $1.87\pm 0.04$ & \nodata & \nodata & $1.63^{+0.12}_{-0.11}$ & 127/77\\
\tableline
  & BKN PL & SMC & $4.02^{+0.62}_{-0.67}$ & $1.58^{+0.02}_{-0.03}$ & $1.30^{+0.19}_{-0.11}$ & 2.08 & $0.50\pm 0.04$ & 101/76\\
 & & LMC & $4.41^{+0.69}_{-0.63}$ & $1.62\pm 0.03$ & $1.30^{+0.16}_{-0.14}$ & 2.12 & $0.72\pm 0.06$ & 99/76\\
 & & MW & $4.78^{+0.75}_{-0.65}$ & $1.67\pm 0.04$ & $1.35^{+0.16}_{-0.17}$ & 2.17 & $1.02^{+0.11}_{-0.10}$ & 102/76\\
\tableline
\end{tabular}
\begin{list}{}{}

     \item $^a$ $\Gamma_2$
  is set to be equal to $\Gamma_1$~+~0.5 in each broken power-law fit, as
  would be expected if the change in index were due to a cooling break.
      \item $^b$ In the fit to the peak epoch, $A_V$ is fixed to the average
      best-fit value found in the same model fits to plateau and shallow stage
      data. The $A_V$ values are given for the observer's frame of reference.

    \end{list}

\caption{Power-law (PL) and broken power-law (BKN PL) fits to the simultaneous
  UVOT and XRT spectra of GRB~061121, for three different dust extinction models: Small
and Large Magellanic Clouds (SMC and LMC) and the Milky Way (MW).
  $\Gamma_{1}$ and  $\Gamma_{2}$ are the photon indices below and above the
  spectral break for the BKN PL models. The data points have not been
  corrected for reddening.}
\label{uvotsed}
\end{center}
\end{table*}


\section{Discussion}
\label{disc}

{\it Swift} triggered on a precursor to GRB~061121 leading to comprehensive broadband
observations of the prompt emission, as well as the later afterglow. We
discuss these here, together
with possible mechanisms involved.

\subsection{Precursor}
Lazzati (2005) found that about
20\% of BATSE bursts showed evidence for gamma-ray emission above
the background between 10 to $\sim$200~s before the main burst, typically with
non-thermal spectra which tended to be softer than the main burst. GRB~060124
(Romano et al.\ 2006) and GRB~061121 show the same behaviour.

Precursor models have been proposed for emission well-separated from, or just
prior to, the main burst. Early emission
occurring only a few seconds before the main burst has been explained by the
fireball interacting with the massive progenitor star -- though the spectrum
of such emission is expected to be thermal (Ramirez-Ruiz et al.\
2002a). 
Lazzati et al.\ (2007) investigated shocks in a cocoon around the main burst; their
model predicts a non-thermal precursor as the jet breaks out of the surface of
the star. A high-pressure cocoon is formed as the sub-relativistic jet head
forces its way out of the star. As the head of the jet breaks through the surface, the energy of
this cocoon is released through a nozzle and can give rise to a
precursor (Ramirez-Ruiz et al.\ 2002a,b). Within the framework of this model,
observers located at viewing angles of 5$^\circ$~$<$~$\theta$~$<$~11$^\circ$ are
expected to see first a relatively bright precursor, then a dark phase with
little emission, followed, when the jet enters the unshocked phase, by a
bright GRB; this is very similar to the light-curve observed for GRB~061121. 
Waxman \& M{\' e}sz{\' a}ros (2003) demonstrate that both a series of thermal X-ray
precursors (becoming progressively shorter and harder) and nonthermal emission
can be produced by an emerging shocked jet, although the nonthermal component
is expected to be in the MeV range. There could also be an accompanying
inverse Compton component, formed by the thermal X-rays being upscattered by
the jet.

The same type of smooth, wide-pulse, low intensity emission as seen in some
precursors, but occurring {\it after} the main emission is also occasionally
seen (e.g., Hakkila \&
Giblin 2004; Nakamura 2000). Hakkila \& Giblin (2004) discuss two examples
where postcursor emission is found to have a longer lag than expected from the
lag-luminosity relation, smoother shape and to
be softer. In the case of the GRB~061121 precursor, the spectrum is, indeed, softer than the main event, and shows a
comparatively smooth profile. The emission does have a longer
lag than the main emission, but it is still  consistent with the lag-luminosity
relation. 
 
There are two expected effects which could lead to
such a difference in lags for separate parts of a single burst: the much lower
luminosity for the precursor (resulting from a much smaller Lorentz factor;
the measured fluence of the precursor is about a factor of 30
smaller than the fluence of the main emission) is
a natural explanation, while the precursor being emitted at a greater
off-axis angle could also have an effect. In this second case, ejecta are
considered to emerge at
different angles with respect to the jet axis; not all of the solid angle of
the jet will be `filled' uniformly.

Such late postcursor emission is
unlikely to be linked to the jet breakout from the stellar surface, and it
may not be sensible to attribute apparently similar phenomena (in
the form of pre- and postcursors) to entirely different processes.

Pre/postcursor emission could be due to
the deceleration of a faster front shell, resulting in
slower shells catching up and colliding with it (Fenimore \& Ramirez-Ruiz
1999; Umeda et al.\ 2005; note, however, that a {\bf faster} shell would be
inconsistent with the precursor having a smaller Lorentz factor as suggested to
explain the lag discrepancy), or late activity of the
central engine. The presence of flares
in about 50\% of {\it Swift} bursts is generally attributed to continuing
activity of the central engine (Burrows et al.\ 2005b; Zhang et al.\
2006a) and the appearance of broken power-laws in the X-ray spectra of both
flares and the prompt emission (Guetta et al.\ 2006; Goad et al.\ 2007) hints
of a common mechanism.

\subsection{Prompt Emission}

The prompt emission mechanism for GRBs is still debated and the origin of
E$_{\rm peak}$ is not fully understood (M{\'e}sz{\' a}ros et al.\ 1994;
  Pilla \& Loeb 1998; Lloyd \& Petrosian 2000; Zhang \& Meszaros
2002; Rees \& M{\'e}sz{\' a}ros 2005; Pe'er et al.
2005). The standard synchrotron model predicts fast
cooling (Ghisellini et al.\ 2000) with a photon
index, $\Gamma$, of 3/2 and (p/2)+1 below and above the peak energy,
  respectively (e.g., Zhang \& M{\'e}sz{\' a}ros 2004). The {\it Konus-Wind} spectral index below E$_{\rm peak}$ is shallower
  than 3/2, which may suggest a slow cooling spectrum with p~$<$~2 [E$_{\rm peak}$ being the cooling
frequency and $\Gamma$~=(p+1)/2] or additional heating. A slow-cooling spectrum can be retained by
assuming that the magnetic fields behind the shock decay significantly in 10$^{4}$--10$^{5}$~cm, so that synchrotron emission happens in small scale magnetic
fields (Pe'er \& Zhang 2006). 

The SED at the peak time (SED~2 in
  Figure~\ref{sed}, discussed below) has a peak
  flux density of around 1~keV, below
which the optical to X-ray spectral slope is 0.11~$\pm$~0.09. This slope
is harder than expected from the standard synchrotron model (which predicts
an index of 1/3). There should, however, be spectral curvature around the break,
which could flatten the index (Lloyd \& Petrosian 2000), so the data could still be consistent with the
synchrotron model. An alternative to synchrotron emission, in the form of
`jitter' radiation is discussed by
Medvedev (2000), though that model predicts an even steeper index of 1 below
the jitter break frequency.


Figures~\ref{joint} and \ref{all-lc} show that all three instruments onboard {\it Swift}
saw the prompt emission around 75~s after the BAT trigger. However, it is
noticeable that most of the emission is in the gamma-ray and X-ray bands, with
the optical showing a relatively small increase in brightness in
comparison. Assuming the observed process is synchrotron, then the prompt emission which {\it is} detected by the UVOT will be the
low-frequency extension of this in the internal shock. No reverse
shock is apparent.

\subsection{Afterglow Emission}
\label{after}

\subsubsection{Broken Power-law Decline Models}
\label{bkn}

The afterglow of GRB~061121 was observed over an even broader energy range
(from radio to X-rays) than the prompt emission, with multi-colour data being
obtained from $\sim$~100--10$^{5}$~s after the trigger.
The X-ray light-curve shows evidence
for substantial curvature at later times (see Figure~\ref{wtimlc}), as has been found for
other {\it Swift} GRBs (e.g., GRBs 050315 -- Vaughan et al.\ 2006; 060614 -- Gehrels
et al.\ 2006b). The standard practice has been to fit such a decay using a series of
power-law segments as a function
of time. An alternative
exponential-to-power-law description of the light-curve is given in
$\S$\ref{rwmodel}.

Nousek et al.\ (2006) and Zhang et al.\ (2006a) have both discussed the  canonical shape that many {\it Swift} afterglows seem to follow: steep to
  plateau to shallow, with some light-curves showing a further steepening. In
  these previous works, the
  extrapolation of the BAT data into the XRT band was incorporated into the
  derivation of the steep decay at the start of the canonical light-curve
  shape. In the case of GRB~061121, the full curve can be seen entirely in
  X-rays, suggesting that the previous extrapolations are reliable. 
 For
  the afterglow of GRB~061121, only data after the end of the main burst have been
  modelled with power-laws. The early steep decline, which might be attributable to
  the curvature effect (Kumar \& Panaitescu 2000; Dermer 2004;
  Fan \& Wei 2005), is not considered here.

\begin{figure*}
\begin{center}
\includegraphics[clip, angle=-90,width=12.0cm]{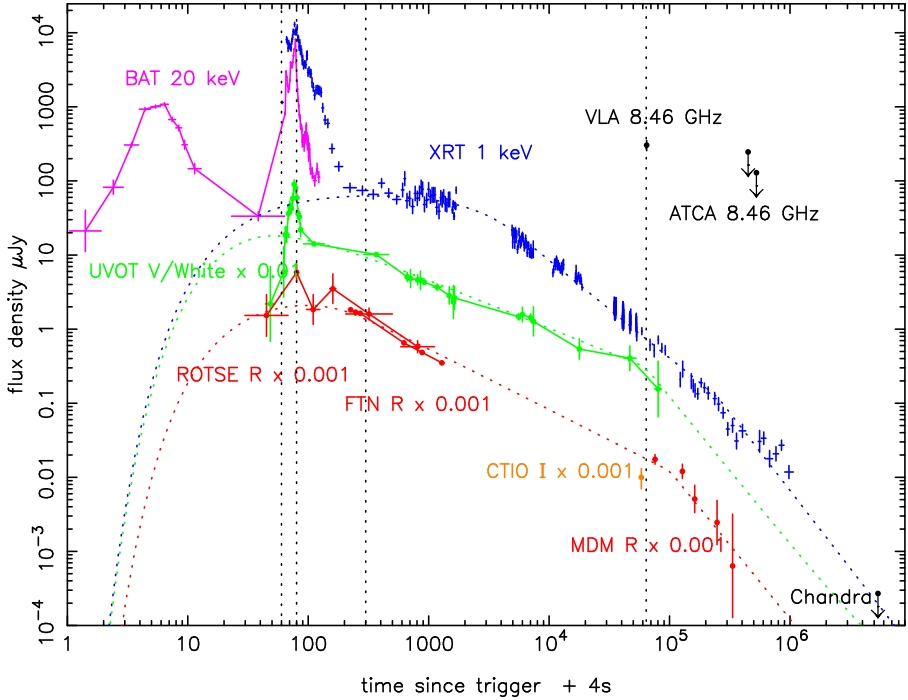}
\caption{Flux density light-curves for the gamma-ray, X-ray, optical and radio
  data obtained for GRB~061121. The vertical dotted lines indicate the times
  used for the SED plots shown in Figure~\ref{sed}, while the curved dotted
  lines show the fit to the X-ray and optical data, including a late-time break, as described in the text.}
\label{all-lc}
\end{center}
\end{figure*}

According to the model proposed in Nousek et al.\ (2006) and Zhang et
al.\ (2006a), the plateau
phase of the light-curve is due to energy injection in the
fireball. The plateau phase of GRB~061121 is consistent with an injection of energy since the luminosity
index, $q$, is negative, which is the requirement for
injection to modify the afterglow (Zhang et al.\ 2006a); the later
two stages both have $q$~$>$~1. 
However, as will be discussed in $\S$\ref{rwmodel}, the plateau and final
transition to the power-law decay are only
visible in the X-ray data for GRB~061121; the start of the final decay is much earlier in the
$V$ and $R$-bands (see Figure~\ref{all-lc}). One might expect that energy injection would affect all the energy bands simultaneously, rather than just the
X-rays.

From the
standard afterglow model computations (e.g., Zhang \& M{\'e}sz{\' a}ros 2004),
we find that none of the closure relations fit the entire dataset
completely: although the shallow phase (after the end of energy injection, between T~+~2.3~ks and T~+~32~ks)
could be consistent with
the evolution of a blast-wave which had already entered the slow cooling regime when
deceleration started [i.e., $\nu > \mathrm{max}(\nu_m,\nu_c)$ where $\nu_c$ is
  the cooling frequency and $\nu_m$ is the synchrotron
  injection frequency; Sari et al.\ 1998; Chevalier \& Li 2000], the steeper part of the decay curve (T~$>$~32~ks) is
not consistent with any of the models. 
This lack of consistency suggests that
a different approach is required.

The change in decay slope between the shallow to steep phases
($\sim$~32~ks) cannot be easily identified with a
jet-break. It certainly seems unlikely that the simplest side-spreading jet model could be applicable, since the post-break decay index ($\alpha$~$\sim$~1.5) is not steep
enough (a post-jet decay has $\alpha$~=~p, where p is the electron index). There is also some indication that the X-ray spectral slope
hardens after the break, whereas no change in spectral signature is expected over a
jet-break. In the case of a non-laterally expanding jet (Panaitescu \&
M{\'e}sz{\' a}ros 1999),
$\alpha$~=~(3$\beta$/2)~+~0.25 [for a homogeneous circumstellar medium (CSM); Panaitescu
et al.\ (2006)], which does, indeed, fit the data after this break:
[1.5~$\times$~(0.9~$\pm$~0.08)]~+~0.25~=~1.6~$\pm$~0.1; the measured $\alpha$ is
  1.53. Such a confined jet has been suggested as an explanation for the
  observed decay in a number of previous bursts (e.g., GRB~990123 -- Kulkarni
  et al.\ 1999; GRB~050525A --
  Blustin et al.\ 2006; GRB~061007 -- Schady et al.\ 2007b). The UVOT data obtained around this time show little evidence for
  a break, whereas jet breaks should occur across all energy bands
  simultaneously. However non-simultaneity could be explained by a
  multi-component outflow, where the X-ray emission is produced within a
  narrow jet, while the optical component comes from a wider jet with lower
  Lorentz factor (Panaitescu \& Kumar 2004; Oates et al.\ 2007). 
There remains the issue, however,
  that $\alpha$ should steepen by 0.75 over a jet break (M{\'e}sz{\' a}ros \&
  Rees 1999), whereas the maximum observed change (within the 90\% errors) is only
  $\Delta\alpha$~$<$~0.61, excluding $\Delta\alpha$~=~0.75 at almost
  3$\sigma$; also, again there should be no spectral evolution across the break.
There is, however, a probable jet break at later times, which will be covered in the next Section. 

Other multi-component models [see, e.g., Oates et al.\ (2007) and references therein]
also fail to explain the data, because of the lack of observed energy
injection (plateau phase) in the optical data.

Panaitescu et al.\ (2006a) discuss chromatic breaks in {\it Swift} light-curves,
and postulate that these could be due to a change in microphysical
parameters within a wind environment. However, this model requires the cooling
frequency to lie between the X-ray and optical bands and, as will be discussed
in $\S$\ref{rwmodel}, this does not seem to be the case here.


\begin{figure}
\begin{center}
\includegraphics[clip, angle=-90,width=8.0cm]{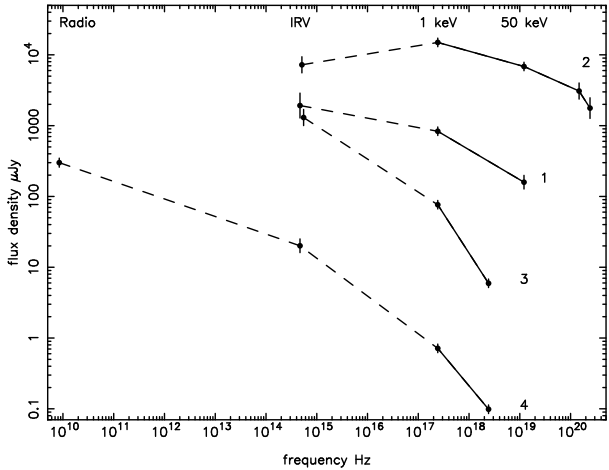}
\caption{SEDs for the four time intervals indicated
  in Figure~\ref{all-lc}. SED~2 (the peak of the burst emission) includes the
  {\it Konus-Wind} data, although these have not been included in
  Figure~\ref{all-lc}. The solid lines represent the power-law fits to the
  BAT, XRT and {\it Konus} data, while the dashed lines join the radio,
  optical and 1~keV points. Spectral evolution over time is clearly seen.}
\label{sed}
\end{center}
\end{figure}


\subsubsection{Exponential-to-power-law Decline Model}
\label{rwmodel}

As first described by O'Brien et al.\ (2006), and further expanded by Willingale
et al.\ (2007), GRB light-curves can be well modelled by one or two
components comprised of an early exponential rise followed by a power-law decay
phase. Of these components, the first represents the prompt gamma-ray emission
and early X-ray decay. The second, when detected, dominates at later times,
forming what we see as the afterglow. These results show that fitting an
intrinsically curved decay with multiple power-law segments runs the risk of
incorrectly identifying temporal breaks (see also Sakamoto et al.\ in prep). In this Section the models of
O'Brien et al.\ (2006) and Willingale et al.\ (2007) are applied to the
multi-band afterglow data of GRB~061121.

Figure~\ref{all-lc} brings together the BAT, XRT, UVOT, FTN and {\it ROTSE} data,
along with further optical and radio points taken from the GCN Circulars
(Halpern et al.\ 2006a,b; Halpern \& Armstrong 2006a,b; Chandra \& Frail 2006;
van der Horst et al.\ 2006a,b) and the upper limit from {\it Chandra}, to form a multi-energy decay plot. The data
have been plotted as `time since trigger + 4~s' in order to include the
precursor on a log time-scale.
The optical
points have all been corrected for extinction using  A$_{\rm V}$~=~1.2 (a combination of the
Galactic value of 0.151 and an estimate of A$_{\rm V}$~$\sim$~1 for the GRB host
galaxy -- see $\S$\ref{sec:uvot-xrt}).

The contribution from the host galaxy reported by
Malesani et al.\ (2006) and Cobb (2006) has been subtracted
from the $V$- and $R$-band flux values. The magnitude of the host in the $V$-band is
22.4, which only changes the last two or three $V$-band
points by a small amount. For the $R$-band we have no direct
measurement, but the last group of MDM exposures
gave an $R$ magnitude of 22.7, corresponding to a flux level of 2.8~$\mu$Jy,
and the flux level is still declining at that epoch
($\sim$~3.3~$\times$~10$^{5}$~s), so an $R$-band flux level of 2.5~$\mu$Jy was
adopted for the host. The error
bars shown on the last few points reflect the large uncertainty
in the galaxy contribution subtracted.

The curved dotted lines in Figure~\ref{all-lc} are the fits to the data
using the exponential-to-power-law model, followed by a break to a steeper
decay around 10$^{5}$~s. These models are parameterised by the power-law
decay, $\alpha$, and T$_{\rm a}$, the time at which
this decay is established. For the
X-ray data, T$_{\rm a,X}$ is found to be 5250$^{+500}_{-460}$~s and
$\alpha_{\rm a,X}$~=~1.32~$\pm$~0.03. Fits were also performed to the $V$- and $R$-band data, yielding:
$\alpha_{\rm a,V}$~=~0.66~$\pm$~0.04 (with T$_{\rm
  a,V}$~=~70$^{+60}_{-70}$~s) and
$\alpha_{\rm a,R}$~=~0.84~$\pm$~0.03 (T$_{\rm a,R}$~=~230$^{+120}_{-230}$~s). 

The non-detection by {\it
  Chandra} almost two months after
the burst shows there must have been a further steepening in the X-ray
regime, and the optical data are not inconsistent with this
finding. Constraining the temporal index after the late break to be
$\alpha$~=~2 (a typical slope for a post-jet-break decay), break times of $\sim$~2.5~$\times$~10$^{5}$,
$\sim$~2.5~$\times$~10$^{4}$ and $\sim$~10$^{5}$~s are estimated for the
X-ray, $V$- and $R$-band respectively; note that the UVOT $V$-band value is
  particularly uncertain, given the small number of data points at late times. Within the uncertainties, these times
  are likely to be consistent, so the turnover could be achromatic, as required
  for a jet break.
From Willingale et al.\ (2007), a jet break might be expected at
$\sim$~100~$\times$~T$_{\rm a,X}$ -- i.e., 5.5~$\times$~10$^{5}$~s, which is
in agreement with these fits.

As can be seen from these numbers and the models plotted in
Figure~\ref{all-lc}, the X-ray data clearly show the transition from the
plateau to the power-law decay, whereas the start of the final decay is much
earlier in the $V$- and $R$-bands. 
The $V$-band decay is also significantly
flatter (by $\alpha$~$\sim$~0.2) than that estimated for the
$R$-band. 
As previously stated, the $V$,
$B$ and $U$ light-curves are all consistent with this slow decay. There have
been few multi-colour optical decay curves obtained
for GRB afterglows, and, of these, the different filters [in the case of GRB~061007 (Schady
  et al. 2007; Mundell et al. 2007) X-ray and gamma-ray data as well as the optical] tend to
  track each other (e.g., Guidorzi et al. 2005; Blustin et al. 2006; de Ugarte Postigo et
  al. 2007). In the case of GRB~061121, we find that the $R$-band
  data are fading more rapidly than the $V$. GRB~060218, which was associated
  with a supernova (e.g., Campana et al. 2006a), shows changes throughout the
  optical spectra, because of a combination of shock break-out and radioactive
  heating of the supernova ejecta. There is a large difference between the
  decays of the blue ($V$, $U$, $B$) and red ($R$) data for GRB~061121, which
  cannot be easily explained by a synchrotron spectrum. Although no supernova
  has been detected in this case, we speculate that some form of pre-supernova thermal
  emission could possibly be affecting the optical data, adding energy into the blue
  end of the spectrum, thus slowing its decline.

After the break in the decays around 10$^{5}$~s, the light-curves
across all bands become more
consistent with one another, although there are only limited data at such a
late time.

The vertical
dotted lines in Figure~\ref{all-lc} show the times of the SEDs plotted in
Figure~\ref{sed}; again, all points were corrected for an extinction of
A$_{\rm V}$~=~1.2, so that
they represent the true SEDs (with the frequency in the
observer's frame). The solid
lines represent actual fits to the X-ray and gamma-ray data, while the dashed
lines just join the separate radio, optical and 1~keV points. 
The times of these SEDs, which clearly show spectral evolution, correspond to (1) before the main BAT peak, 56~s after trigger;
(2) at the BAT peak, 76~s after trigger; (3) just after the start of the
plateau, 300~s after the trigger; (4) in the main decay at 65~ks
(chosen because radio measurements were taken at this time). SEDs~3 and 4 do
not contain any BAT or {\it Konus} data, since the gamma-ray flux had decayed
by this point; the highest energy point in these corresponds
to the maximum energy (10~keV) of the X-ray fits.

Table~\ref{uvotsed} demonstrates that
the optical and X-ray spectra during the peak emission are best
fitted with a broken power-law model, with the break energy at the very low
energy end of the X-ray bandpass. SED~2 in Figure~\ref{sed} shows that this
spectral break corresponds to the peak frequency in a flux density plot
($\beta_{1}$ is less than zero in this case). Only during SED~2 is the optical
flux density lower than that of the higher energy data. Figure~\ref{joint}
also shows that the optical emission is less strong than the X-ray and
gamma-ray data during the main
burst.

\begin{table*}
\begin{center}
\begin{tabular}{lcccccc}
\tableline
\tableline
GRB models  & $\alpha(\beta)$ & $\alpha(\beta_{\mathrm{a,X}})^a$ &  
$\alpha^{b}_{\mathrm{a,X}}$ & $\alpha(\beta_{\mathrm{opt}})^a$ &
\multicolumn{2}{c}{$\alpha^{b}_{\mathrm{opt}}$}\\
& & & & & $V$-band & $R$-band\\
\tableline
CSM SC$^c$ ($\nu_m< \nu<\nu_c$) &  $\frac{3\beta}{2}$ & 1.49~$\pm$~0.10 & 1.32~$\pm$~0.03
 & 0.80~$\pm$~0.09 & 0.66~$\pm$~0.04 & 0.84~$\pm$~0.03 \\
Wind SC$^c$ ($\nu_m< \nu<\nu_c$) &  $\frac{3\beta+1}{2}$ & 
 1.99~$\pm$~0.10 & &  1.30~$\pm$~0.09   \\
CSM or Wind SC$^c$ \& FC$^d$  & $\frac{3\beta-1}{2}$
& 0.99~$\pm$~0.10  & & 0.30~$\pm$~0.09 \\
($\nu > \mathrm{max}(\nu_c,\nu_m)$) & & & \\
\tableline
\end{tabular}
    \begin{list}{}{}
     \item $^a$ Decay calculated from the measured spectral index
     \item $^b$ Observed power-law decay index.
      \item $^c$ Slow cooling.
      \item $^d$ Fast cooling.
    \end{list}
\caption{Closure relations for
  exponential-plus-power-law model fits to the X-ray data
  ($\beta_{\mathrm{a,X}}=0.99\pm 0.07$) and the optical-to-X-ray band
  ($\beta_{\mathrm{opt}}$~=~0.53~$\pm$~0.06) from the
  time of SED~4 (65~ks after the burst).}
\label{closure}
\end{center}
\end{table*}

Table~\ref{closure} shows the values of $\alpha$ for the X-ray and
optical decays (i.e., before and after the break) in SED~4, at 65~ks, with their corresponding spectral indices. 
For the initial stages of the power-law decay (T$_a$~$<$~t~$<$~65000~s) the evolution of the
afterglow SED and the coupling between the temporal and
spectral indices are not completely consistent with the standard
model: although the $R$-band decay, with $\alpha_{\rm a,R}$~=~0.84~$\pm$~0.03,
is in good agreement with the homogeneous CSM model below the cooling break, the  X-ray and
$V$-band flux decays are slower
than expected from the measured spectral indices; they are in best agreement
with the same constant density model below $\nu_c$, however. 


The point at which the power-law decay dominates the exponential in the optical bands is noticeably
earlier than in the X-ray ($<$ few hundred seconds, rather than $\sim$~5000~s)
and, as mentioned above, the decay indices are significantly
different for all three (X-ray, $V$ and $R$) bands (see
Figure~\ref{all-lc}). At the time of SED~3, the X-ray data are not decaying
(i.e., this is during the plateau), yet both the $V$ and $R$-band data
have already entered the power-law decline phase. 
The $R$-band is decaying faster
than the $V$-band, so the spectral index through the optical range is becoming
harder. 
The X-ray spectral index
shows a similar hardening trend (see Table~\ref{xrtspec}), so the SED measured from optical
to 10~keV is gradually getting harder. Such spectral hardening from the plateau to the final decay is a feature of
many X-ray afterglows (Willingale et al.\ 2007).

This slow hardening of the broadband spectrum with time could be a
signature of synchrotron self-Compton emission (Sari \& Esin 2001;
Panaitescu \& Kumar 2000).
The strength of the self-Compton component in the afterglow depends on
the flux of low energy photons (radio-optical) and the electron density
in the shock. Using the formulation in Sari \& Esin (2001) the density
required is given by

\begin{equation}
n_{1}=3\times10^{9}\left(\frac{f^{IC}_{max}}{f^{sync}_{max}}\right)^{4/3}
(E_{52}t_{day})^{-1/3} cm^{-3}
\end{equation}

where $f^{\rm IC}_{\rm max}/f^{\rm sync}_{\rm max}$ is the ratio of the peak
flux of the
seed synchrotron spectrum (i.e., the source of low energy photons) and the peak
flux of the self-Compton emission; $E_{52}$ is the isotropic burst
energy in units of $10^{52}$~erg; $t_{day}$ is the time in days after the burst
(which determines the distance through the CSM swept up by the external
shock). From Figure~\ref{sed} (SEDs 1, 3 and 4) we see that $f^{\rm IC}_{\rm
  max}/f^{\rm sync}_{\rm max}$~$\sim$~0.001
if the X-ray flux has a significant contribution from a self-Compton
component at $t_{\rm day}$~=~0.75. A value of $E_{52}$~=~30 gives $n_{1}\approx10^5$~cm$^{-3}$. Even assuming the emission at 0.75~days is
not dominated by the self-Comptonisation, and so taking the
$f^{\rm IC}_{\rm max}/f^{\rm sync}_{\rm max}$ ratio to be a factor of ten smaller, the density
would be $\sim$~5~$\times$~10$^3$~cm$^{-3}$, which is still high. It seems unlikely that self-Compton emission is the cause of the spectral
hardening of the SED unless the CSM density encountered by the external
shock is extremely large. However, there have been suggestions that GRBs may form in
molecular clouds (Galama \& Wijers 2001; Campana et al.\ 2006b,c), which have densities of 10$^{4}$ or more particles per cubic
centimetre in the cores (Miyazaki \& Tsuboi 1999; Wilson et al.\ 1999). Typically one might expect greater reddening than is found
here (Table~\ref{uvotsed}), though Waxman \& Draine (2000) discuss the
possibility of dust destruction.


The spectrum will be redshifted as the jet slows down, so the optical and
X-ray spectral indices should, if anything, become softer -- the opposite of what
is seen here. Although spectral hardening with time is suggested from the data, it is not be easily explained by current models.

Whether or not there is a Comptonised component, the later SEDs clearly indicate that
there is a break in the spectrum
somewhere between the optical and the X-ray; this is also shown by the
fits in Table~\ref{uvotsed}, where the UVOT--XRT spectra are better fitted
with broken power-laws, with E$_{\rm break}$ towards the low energy end of the
X-ray bandpass. Since both the optical and X-ray
bands appear to be below the cooling frequency, from the closure relations
given in Table~\ref{closure}, this change in slope cannot be identified with a
cooling break; its origin remains unclear.


The redshift of $z=1.314$ and the isotropic energy  of
E$_{\rm iso}$~$\sim$~3~$\times$~10$^{53}$~erg ($\S$\ref{sec:bat-xrt}) can be
used to place constraints
on the jet opening angle. From Sari et al.\ (1999), and
assuming that the jet break occurs at $T_0+2\times 10^5$\,s, we have $ \theta_j \sim 4^\circ
\left(\frac{\eta_{\gamma}}{0.2}\right)^{1/8}~
\left(\frac{n}{0.1}\right)^{1/8}$ where $n$
and $\eta_{\gamma}$ are the density of the CSM and the efficiency of
the fireball in converting the energy in the ejecta into gamma-rays. Taking
$\eta_{\gamma}$~=~0.2 and n~=~3~cm$^{-3}$ (following Ghirlanda et al.\ 2004), this gives E$_{\gamma}$~$\sim$~1.7~$\times$~10$^{51}$~erg
for the beaming-corrected gamma-ray energy released, which is within
the range previously determined (e.g., Frail et al.\ 2001) and consistent with
the Ghirlanda relationship (Ghirlanda et al.\ 2004).

\section{Summary and conclusions}
\label{conc}

{\it Swift} triggered on a precursor to GRB~061121, leading to unprecedented
coverage of the prompt emission by all three instruments onboard, with  
the gamma-ray, X-ray and optical/UV
bands all tracking the main peak of the burst. GRB~061121 is the instantaneously brightest long
{\it Swift} burst detected thus far, both in gamma-ray and X-rays.
The precursor and main burst show spectral lags of different lengths, though
both are consistent with the lag-luminosity relation for long GRBs (Gehrels et
al.\ 2006b). 

The SED of the prompt emission,
stretching from 1~eV to 1~MeV shows a peak flux density at around
1~keV and is harder than the standard model predicts.
There is definite curvature in the spectra, with the prompt optical-to-X-ray
spectrum being better fitted by a broken
power-law, similar to results found for fitting X-ray flares (e.g.,
Guetta et al.\ 2006; Goad et al.\ 2007).

The afterglow component, in both the optical and X-ray, starts early
on -- before, or during, the main burst peak (see also O'Brien et al. 2006;
Willingale et al. 2007; Zhang et al. 2006b). The broadband SEDs reveal gradual
spectral hardening as the afterglow evolves, both within the X-ray regime
($\Gamma$ flattening from $\sim$~2.05 to $\sim$~1.87) and
between the $V$- and $R$-band optical data ($\alpha_{\rm
V}$~$\sim$~0.66 compared with $\alpha_{\rm R}$~$\sim$~0.84). Self-Comptonisation could explain
the hardening, although a molecular-cloud-core density would be required.
A probable jet-break occurs around T$_{\rm 0}$~+~2~$\times$~10$^{5}$~s, shown
by a late-time non-detection by {\it Chandra}. Before this break, the X-ray and
$V$-band decays are too
slow to be 
readily explained by the standard models.


This extremely well-sampled burst shows clearly that there remains much work
to be done in the field of GRB models. A single, unified model for all GRB emission observed should be the ultimate goal.

\section{Acknowledgments}

The authors gratefully acknowledge support for this work at the University of
Leicester by PPARC, at PSU
by NASA and in Italy by funding from ASI. This work is partly based on observations 
with the {\it Konus-Wind} experiment (supported by the Russian Space Agency
contract and RFBR grant 06-02-16070) and on data
obtained with {\it XMM-Newton}, an
ESA science mission
 with  instruments and contributions directly funded by
 ESA Member States and NASA.
We thank the Liverpool GRB group at ARI, Liverpool John Moores University,
in particular C.J. Mottram, D. Carter, R.J. Smith and A. Gomboc for their
assistance with the FTN data acquisition and interpretation. The Faulkes Telescopes are operated by the Las Cumbres Observatory
Global Telescope Network. 
We also thank J.E. Hill and 
A.F. Abbey for discussions and help with the PuPD data, P. Curran for assistance
with the UVOT-XRT SED creation, D. Grupe for the {\it Chandra} upper limit and B. Cobb and D. Malesani for
information regarding the magnitude of the host galaxy. Thanks as well to C.
Akerlof, E. Rykoff, A. Phillips and M.C.B. Ashley from the ROTSE team.


\end{document}